\documentclass[aps,pra,preprintnumbers,superscriptaddress,10pt,twocolumn,floatfix]{revtex4-2}

\usepackage{xr}
\usepackage[version=3]{mhchem}
\usepackage{graphicx}
\usepackage{mathtools}
\usepackage{mathrsfs}
\usepackage{txfonts}
\usepackage{upgreek}
\usepackage[colorlinks=true,citecolor=blue,urlcolor=red,verbose]{hyperref}
\usepackage{lipsum}
\usepackage{tensor}
\usepackage[utf8]{inputenc}
\usepackage[T1]{fontenc}
\usepackage[exponent-product = \cdot]{siunitx}
\usepackage{cleveref}           
\usepackage{natbib}
\usepackage{xcolor}
\usepackage{array,multirow,graphicx}
\usepackage{booktabs}
\usepackage{color}
\usepackage[mathscr]{eucal}

\DeclareDocumentCommand\vectorbold{ s m }{\IfBooleanTF{#1}{\boldsymbol{#2}}{\mathbf{#2}}}
\DeclareDocumentCommand\vb{}{\vectorbold}

\newcommand{\bolds}[1]{\boldsymbol{#1}}
\newcommand{\bmc}[1]{\boldsymbol{\mathcal{#1}}}

\newcommand{\p}{\partial}
\newcommand{\bn}{\bolds{\nabla}}
\newcommand{\tn}{\textnormal{N}}

\DeclareMathOperator{\idb}{\vb{Id}}

\DeclareMathOperator{\grada}{grad}

\newcommand{\mass}{M}
\newcommand{\massdens}{\upvarrho}
\newcommand{\chargedens}{\varrho}
\newcommand{\conc}{c}
\newcommand{\entro}{s}
\newcommand{\vel}{\vb{v}}

\newcommand{\freedens}{\upvarphi_\textnormal{H}} 

\newcommand{\freedensdot}{\dot{\upvarphi}_{\textnormal{H}}}

\newcommand{\stress}{\bolds{\sigmaup}}

\newcommand{\heatcond}{\gammaup}
\newcommand{\suszept}{\upchi}
\newcommand{\elcond}{\kappaup}
\newcommand{\bulkmodalt}{\mathcal{K}}
\newcommand{\pmv}{\upnu}
\newcommand{\valence}{z}
\newcommand{\effval}{\tilde{\valence}}
\newcommand{\trans}{t}
\newcommand{\diffusioncomp}{\mathcal{D}}
\newcommand{\diffusion}{\bmc{D}}

\newcommand{\diel}{\varepsilon_0}
\newcommand{\dielrel}{\varepsilon_{\ce{R}}}

\newcommand{\elf}{\vb{E}}
\newcommand{\magf}{\vb{B}}
\newcommand{\delf}{\vb{D}}
\newcommand{\hmagf}{\vb{H}}
\newcommand{\emf}{\bolds{\mathcal{E}}}
\newcommand{\mmf}{\bmc{H}}

\newcommand{\Jflux}{\bmc{J}}

\newcommand{\Nflux}{\vb{N}}
\newcommand{\sflux}{\bolds{\xi}_\entro}

\newcommand{\elpot}{\Phi}
\newcommand{\chempot}{\upmu}
\newcommand{\effpot}{\tilde{\chempot}}
\newcommand{\effeffpot}{\tilde{\effpot}}
\newcommand{\tempr}{T} 

\newcommand{\supi}{SI}
\newcommand{\znoacion}{\ce{[Zn(OAc)_3]^-}}
\newcommand{\oacion}{\ce{OAc^{-}}}
\newcommand{\chion}{\ce{Ch^{+}}}
\newcommand{\choac}{\ce{[Ch]OAc}}
\newcommand{\znoac}{\ce{Zn(OAc)_2}}

\newcommand{\STAB}[1]{\begin{tabular}{@{}c@{}}#1\end{tabular}}

\begin{document}
\sisetup{tight-spacing=true}
\setcitestyle{super}

\allowdisplaybreaks

\title{Theory of Transport in Highly Concentrated Electrolytes}

\author{Max Schammer}
\affiliation{German Aerospace Center, Pfaffenwaldring 38-40, 70569 Stuttgart,
  Germany} 
\affiliation{Helmholtz Institute Ulm, Helmholtzstra{\ss}e 11, 89081 Ulm,
  Germany}  

\author{Birger Horstmann}
\email{birger.horstmann@dlr.de}
\affiliation{German Aerospace Center, Pfaffenwaldring 38-40, 70569
  Stuttgart, Germany}
\affiliation{Helmholtz Institute Ulm, Helmholtzstra{\ss}e 11, 89081
  Ulm, Germany} 
\affiliation{Universit\"at Ulm, Albert-Einstein-Allee 47, 89081 Ulm,
  Germany}

\author{Arnulf Latz}
\email{arnulf.latz@dlr.de}
\affiliation{German Aerospace Center, Pfaffenwaldring 38-40, 70569
  Stuttgart, Germany}
\affiliation{Helmholtz Institute Ulm, 
  Helmholtzstra{\ss}e 11, 89081 Ulm, Germany} 
\affiliation{Universit\"at Ulm,
  Albert-Einstein-Allee 47, 89081 Ulm, Germany} 

\begin{abstract}

  Ionic liquids are promising candidates for novel electrolytes as they
  possess large electrochemical and thermodynamic stability and offer a high
  degree of tunability. As purely-ionic electrolyte without neutral solvent
  they exhibit characteristic structures near electrified interfaces and in
  the bulk, both being described theoretically via separate frameworks and
  methodologies. We present a holistic continuum theory applying to both
  regions. This transport theory for pure ionic liquids and ionic
  liquids-mixtures allows the systematic description of the electrolyte
  evolution. In particular, dynamic bulk-transport effects and interfacial
  structures can be studied. The theory is thermodynamically consistent and
  describes multi-component solutions (ionic liquids, highly concentrated
  electrolytes, water-in-salt electrolytes). Here, we give a detailed
  derivation of the theory and focus on bulk transport processes of ionic
  liquids as appearing in electrochemical cells. In addition, we validate our
  framework for a zinc-ion battery based on a mixture of ionic-liquid and water as electrolyte.

\end{abstract}

\maketitle

\section*{Introduction}
\label{sec:introduction}
The political and social demand for ecologically friendly, low cost,
and rechargeable batteries with high energy densities is an increasing
research stimulus for the improvement of materials, and the
development of novel battery-components.\cite{Crabtree2015}
Electrolytes play a key role for the performance of electrochemical
systems.\cite{armand2008building} Common types of electrolytes include
aqueous electrolytes,\cite{Clark2020} organic
electrolytes,\cite{Hein2020} solid electrolytes,\cite{Braun2015} and
recently, ionic liquids (ILs).\cite{C3EE42099J}
ILs have attracted attention in the context of many technologies,
including energy management, electrodeposition, bioscience, and
biomechanics.\cite{endres_il_future} A large number of ILs can be
mixed from various cations and anions.\cite{Plechkova2008} This allows
to tailor-cut them into task specific designer electrolytes. Thus,
many ILs exhibit a variety of beneficial properties which makes
them promising candidates for novel
electrolytes.\cite{glasstransitionofils} Their favorable
properties include large electrochemical windows, low flammability, or
low vapor pressures. Thus, they can be compatible with high voltage
electrodes, offer intrinsic safety, or be stable towards
air.\cite{978-1-84755-161-0} Furthermore, some ILs suppress
dendritic growth during metal deposition.\cite{endres_dendrites}
Depending on their composition, many ILs are environmentally
friendly.\cite{eco_friendly_ils}

\begin{figure*}[!tb]
  \includegraphics[width=0.9\textwidth]{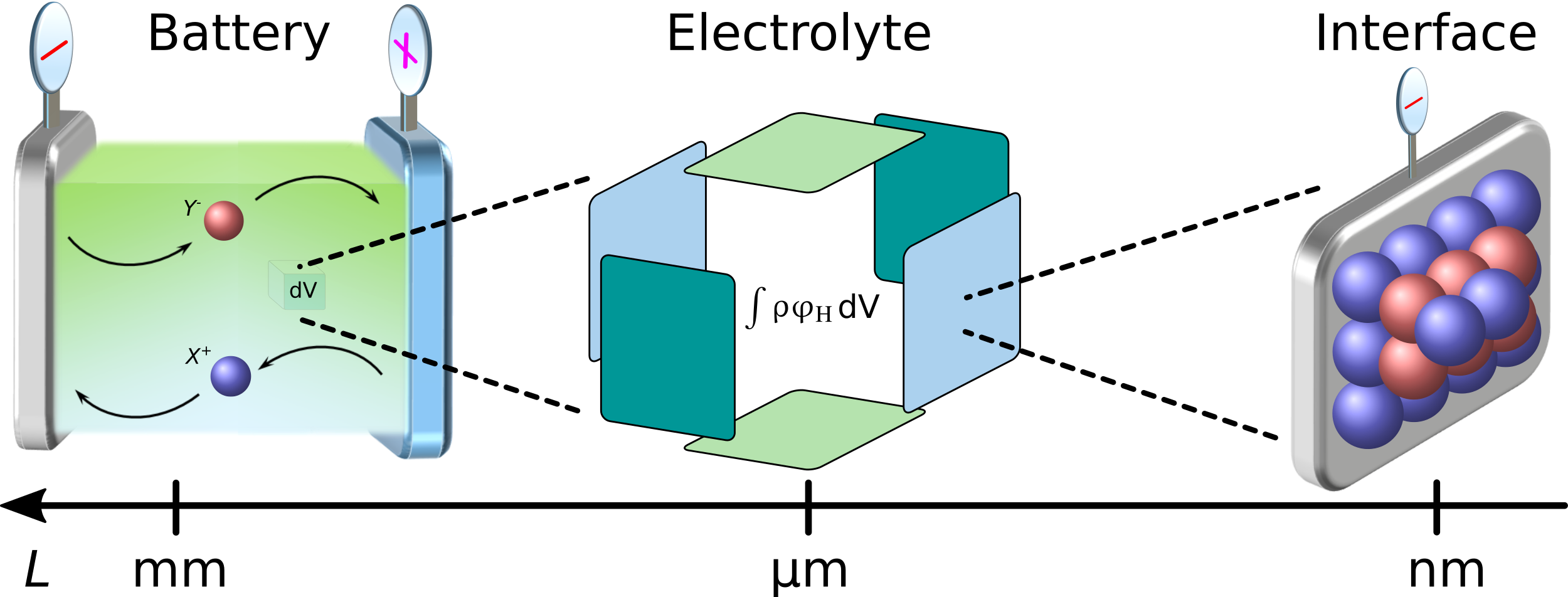}
  \caption{Scheme of our holistic framework, which captures
    length-scales from battery-cells to particle interactions. Thus,
    it describes macroscopic phenomena, like discharging/charging of a
    battery, mesoscopic effects, like specific
    electrolyte-dynamics, but also interfacial effects occurring at
    microscopic scales, like crowding and overscreening.}
  \label{fig:scheme_framework}
\end{figure*}

Current theoretical studies concentrate on the bulk,\cite{Hayes2015}
\textit{or} on interfacial structures.\cite{Fedorov2014} Unifying
approaches describing both scales are rare, but deliver remarkable
results.\cite{doi:10.1021/acs.jpclett.6b00370,
  doi:10.1021/acs.jpclett.7b03048, doi:10.1021/acs.chemmater.5b00780,
  PhysRevE.95.060201, PhysRevE.95.060201} However, a complete unified
approach for the dynamic description of ILs, and multi-component
IL-mixtures, at bulk and interface, is still missing in the
literature. In this article, we present such a unified framework based
on the concepts of rational thermodynamics (RT) which we previously
applied to electrolytes with neutral
solvents.\cite{arnulfbs2,arnulfbs} RT provides a thermodynamically
consistent framework to model a great variety of non-equilibrium
systems. Here, we use RT to derive a continuum transport theory for
strongly correlated electrolytes. In this work, we focus on
bulk-transport. In a previous publication,\cite{C7CP08243F} we showed
how to supplement the theory with hardcore interactions and describe
the interfacial behavior of ILs. We illustrate our holistic framework
in \cref{fig:scheme_framework}.

Molecular/atomistic studies of ILs are based on molecular dynamics
(MD) simulations and (classical) density functional theory (DFT)
simulations. DFT resolves microscopic ion-properties, delivers
detailed insights into the molecular arrangement in the
electrochemical double layer (EDL)\cite{Wu2011} and describes bulk
properties like ion-pair formation\cite{Holloczki2014} or small-scale
ion-diffusion.\cite{Jiang2014} MD simulations resolve the complete
molecular arrangement and describe the evolution of the
nano-structured bulk-landscape of ILs, dependent on external agents
like temperature, electric fields and
pressure.\cite{Margulis2004,CanongiaLopes2006,Yeganegi2012,Li2013,Sharma2016}
However, DFT/MD simulations are limited due to their computational
costs; simulations at length-scales above the nano-meter scale are
hardly accessible by these atomistic methods.

Continuum theories provide a complementary methodology for dynamic
transport simulations of larger systems. Recently, Bazant et
al. proposed a phenomenological mean-field-theory for binary ILs at
electrified interfaces.\cite{bazant} Using a generalized
Landau-Ginzburg functional, the authors show that higher gradients in
the electric potential lead to quasi-crystalline structures near
electrified interfaces. This finding is confirmed by our theory (see
comments below \cref{eq:13}).\cite{C7CP08243F} Yochelis et
al. rationalized this approach, and extended it to bulk phenomena and
ternary ILs.\cite{doi:10.1021/acs.jpclett.6b00370,
  doi:10.1021/acs.chemmater.5b00780, doi:10.1021/acs.jpclett.7b03048,
  PhysRevE.95.060201} Their theoretical work allows fast dynamic
simulations but cannot resolve detailed transport processes.

Dilute and concentrated electrolytes for lithium-ion batteries drive
the development of continuum transport theories. \cite{Newman2004}
Recently, significant effort was put in the rationalization of
consistent theories for neutral-solvent-based electrolytes with high
amount of salt.\cite{arnulfbs2,arnulfbs,Dreyer2018,guhlke_phd}
Although neat ILs constitute the extreme limit of concentrated
electrolytes, where the neutral solvent vanishes, such transport
theories for lithium ion batteries cannot be generalized to
solvent-free electrolytes. This is because ILs exhibit some
exceptional behaviour. For example, ILs form characteristic
quasi-crystalline structures near electrified interfaces.  These
extend over a couple of nanometers, which is not observed in regular
concentrated electrolytes.\cite{doi:10.1021/jp200544b} Our transport
theory for ILs, however, also describes standard
electrolytes,\cite{Single2019} as the IL-effects decrease under
water-di\-lution,\cite{doi:10.1021/acs.jpcc.6b02549} or minor additive
salts.\cite{C7CP08243F}

Transport theories for highly concentrated electrolytes should contain
a condition which prevents ``Coulomb collapse'' of the
ions.\cite{hansen2006theory} The prevalent electrostatic attraction
can be counteracted by vo\-lu\-me\-tric constraints (``mean steric
effects'') or repulsive particle interactions (``atomistic volumetric
exclusion''). Here, we impose a mean-volume constraint for
multicomponent-incompressibility. This results in a threshold for
local ion-concentration due to finite volumes. Near electrified
interfaces, the concentration-threshold leads to crowding
effects.\cite{C7CP08243F} (Imposing repulsive interactions implies
microscopic effects of excluded volume, which lead to
overscreening.\cite{C7CP08243F})

The momentum equation is explicitly used in our derivation. Thereby,
we capture the coupling of electric and mechanical stresses in the
chemical potentials. This has significant consequences in confined
geometries, \emph{e.g.} electrochemical double layers, where
volumetric constraints are in strong competition with Coulomb
forces.\cite{C7CP08243F} Electrolyte momentum is given by the
center-of-mass convection velocity. Thus, the evolution of convection
determines the momentum equation, whereas its variation, expressed by
the rate-of-strain, is mandatory for the correct electro-mechanical
couplings in the stress tensor. Furthermore, convection is a
significant transport-mechanism in solvent-free electrolytes. We take
account for these phenomena, derive a convection-equation and consider
dissipative, viscous stress in the momentum equation.

We use ILs as validation-template, exemplifying extremely correlated
electrolytes. Nevertheless, our theory applies to any liquid
multi-component electrolyte, composed of arbitrarily charged (also,
neutral) species. Thus it exhibits a very general structure. Our scope
of electrolytes includes ILs, IL-salt mixtures, concentrated
electrolytes (including aqueous electrolytes) and novel
``water-in-salt'' electrolytes.\cite{suo2015water} In particular, for
many applications, neat ILs or solutions with high
salt-concentrations are supplemented by additives (\emph{e.g.}{},
water,\cite{doi:10.1021/acsami.6b01592} organic
solvents,\cite{seddon2000influence} or salts\cite{Liu2017}) to improve
their performance as electrolytes. The transport theory presented in
this work can be tailor-cut to any such heterogeneous mixture.

This article is structured into two main parts. In the first part, we
derive our transport theory, which we validate for a zinc-ion battery
in the second part. Finally, we discuss the novel aspects of our
transport theory in relation to previous works.
  
\section*{Transport Theory}
\label{sec:theory}
Our transport theory relies on non-equilibrium thermodynamics,
supplemented with elements of electromagnetic theory, and
mechanics.\cite{kovetz,Dreyer2018,grootmazur,lebon} We split this
theory-chapter into two sections. In the first section General
Transport Theory, we derive the universal framework, which applies to
a wide class of systems. % Here, rational thermodynamics, as established
% by Coleman and Noll, guides our derivation of the equations of motion,
% and ensures consistency with the second axiom of
% thermodynamics.\cite{Coleman1963}
We highlight the logical steps of this derivation in four
subsections. \emph{First}, we derive the entropy production rate,
\cref{eq:60new}, from the fundamental assumption of conservation of
mass, charge, and energy, by formulating the coupled balance equations
for momentum, energy and entropy. We use symmetry arguments to
guarantee that our model is independent from the state of the
observer.\cite{Muller1972,muller2007history} Thus, the set of
variables is restricted to so-called objective quantities (see
\cref*{SIsec:orth-transf}).\cite{liu2002continuum} Our method applies
to a large class of materials and leaves a broad tunability for
distinct physical systems. \emph{Second}, we perform our first
modeling choices and select a variable-set for strongly correlated
electrolytes, \cref{eq:67}. Then, we evaluate the entropy production
for a general free energy.  In the next two steps, we use a linear
Onsager-Ansatz to close the system of differential equations. In the
\emph{third} subsection, we couple the thermodynamic fluxes and
forces, \cref{eq:27}, which ensures a positive entropy
production. \emph{Fourth}, we assume a linear relationship between
viscosity and the velocity-gradients,
\cref{eq:const_eq_viscosity}. This determines the model-dependent
stress-tensor, \cref{eq:mod_stress}.

In the second section of this theory-chapter, we specify our framework
and model the free energy density for ILs, \cref{eq:4}. This casts the
specific properties of the medium into the formalism. Here, these
properties comprise liquid state, polarizability, temperature
dependence, viscosity, and multi-component structure. We further
assume incompressibility, and, in the case of bulk transport,
electroneutrality. These strong correlations allow to reduce the
minimal set of independent variables, \cref{eq:8,eq:2}. Finally, we
state the equations of motion,
\cref{eq:13,eq:49,eq:49b,eq:49a,eq:49c,eq:64}.

\subsection*{General Transport Theory}
\label{sec:non-equil-therm}

\subsubsection*{Second Law of Thermodynamics: Entropy}

As introduction to our derivation, we present the general structure of
a transport theory based on the concepts of non-equilibrium
thermodynamics. Our model is implemented as continuum theory for the
fundamental physical quantities mass, charge, momentum, energy, and
entropy, described by local volume-specific field densities
$\psi_A(\vb{x})$ at position $\vb{x}=(x_1,x_2,x_3)$. We refer to
physical quantities with capital letters $A,B,C,\ldots$, to spatial
dimensions with lowercase letters $i,j,k,\ldots$, and to dissolved
species with Greek letters $\alpha,\beta,\gamma,\ldots$. The time
evolution of $\psi_A(\vb{x})$ is governed by continuity
equations,\cite{slattery1972momentum}
\begin{align}
  \label{eq:law_balance_template_partial}
  \p_t \psi_A
  &= \frac{\p \psi_A}{\p t}
    =  - \bn  \left( \psi_A \otimes
    \vel + \bolds{\xi}_A  \right)  +  r_A, \\ 
  \label{eq:law_balance_template}
  \dot{\psi}_A
  &=
    \frac{\text{d} \psi_A}{\text{d}t}
    = - \psi_A\left(\bn\cdot \vel\right) - \bn
    \bolds{\xi}_A   +r_A.  
\end{align}
The partial time derivative $\p_t \psi_A$ refers to temporal changes
in the laboratory system, and the total time derivative $\dot{\psi}_A$
describes temporal changes relative to a co-moving observer at
velocity $\vel$. Both are related by a convection term,
$\dot{\psi}_A = \p_t \psi_A + (\vel\cdot\bn) \psi_A$.\cite{kovetz} The
convection velocity $\vel$ is defined in the fixed laboratory-frame as
center-of-mass average,
$\massdens\vel=\sum_{\alpha=1}^\tn \massdens_\alpha \vel_\alpha$,
describing bulk electrolyte-momentum per unit mass.

We derive expressions for the production rates $r_A$ and the
non-convective flux densities $\bolds{\xi}_A$ for each field variable
$\psi_A$ in the remainder of this section.

The total mass density
$\massdens=\sum_{\alpha=1}^\tn \massdens_\alpha$ is conserved in the
bulk electrolyte, $\sum_{\alpha=1}^\tn\mass_\alpha r_\alpha=0$, and it
evolves due to convection only,
\begin{equation}
\label{eq:totmassdens}
\p_t\massdens = -\bn (\massdens \vel)
\hspace{3mm}\text{or}\hspace{3mm} \dot{\massdens} = -\massdens(\bn
\vel).  
\end{equation}
This special role of the total mass density is used to express the
continuity equation \ref{eq:law_balance_template} in a simpler form
\begin{equation}
  \massdens \dot{\widetilde{\psi}}_A
  =   - \bn \bolds{\xi}_A   +  r_A
\end{equation}
for the mass-specific field density
$\widetilde{\psi}_A=\psi_A/\massdens$. Individual concentrations
$\conc_\alpha$, however, are affected by non-convective fluxes,
\begin{align}
  \label{eq:concconserv}
  \partial_t \conc_\alpha
  &=
    -  \bn  \left(    \conc_\alpha    \vel    +     \Nflux_\alpha
    \right)  ,
  \\
  \dot\conc_\alpha
  &=  -  \conc_\alpha\bn\vel - \bn  \cdot \Nflux_\alpha ,
\end{align}
where $\Nflux_\alpha = \conc_\alpha (\vel_\alpha - \vel)$ are the flux
densities in the center-of-mass system. We note that these flux
densities are related,
$\sum_{\alpha=1}^\tn \mass_\alpha \Nflux_\alpha =0$.

As (total) charge
$\chargedens = F \sum_{\alpha=1}^\tn \valence_\alpha \conc_\alpha$ is
conserved, we state the balance equation
\begin{align}
  \label{eq:chargecons}
  \p_t   \chargedens  =  - \bn  \left(  \chargedens \vel  + \Jflux  \right)
  + \sum_{\alpha=1}^\tn   F\valence_\alpha   r_\alpha,
\end{align}
where $\Jflux = F \sum_{\alpha=1}^\tn z_\alpha \Nflux_\alpha$ is the
current density in the center-of-mass system.

Total momentum density $\massdens \vb{g}$ comprises kinetic and
electromagnetic contributions. The correct form of the electromagnetic
contribution is under debate in the literature, and various forms have
been suggested.\cite{Henjes1993a} Here, we set
$\massdens \vb{g} = \massdens\vel + \delf\wedge \magf$, using the
Minkowski-momentum.\cite{Medina2014} We use $\delf$ and $\magf$ as a
Galilei-invariant description of the electromagnetic field, which is
consistent with our choice of variables for the free energy density
(see \cref{eq:67}). We formulate momentum conservation using Euler's
first law of mechanics,
\begin{equation}
  \label{eq:7}
  \massdens \dot{\vb{g}}   =   \bn\stress  + \massdens  \vb{b}.
\end{equation}
Thus, $\massdens \vb{g}$ changes due to body-forces
$\massdens \vb{b}$ acting upon the system, \textit{e.g.} gravity, and
due to anisotropic surface-forces described with the Cauchy-stress
tensor $\stress$. Here, the tensor-gradient is defined as
$(\bn \stress)_i = \sum_{j=1}^{3} \p_j\sigmaup_{ij}$. We neglect
molecular orientations (internal spin) in the isotropic bulk liquid,
which implies a symmetric stress tensor,
$\stress=\stress^{T}$.\cite{mueller1985thermodynamics} In the absence
of of gravitation, total momentum is conserved,
$\massdens\vb{b}=0$. Below, we couple energy and momentum balances via
body-forces $\vb{b}$. Thus, our derivation proceeds without an
explicit model for body-forces.\cite{Reich2018}

The evolution of the energy density $\upvarepsilon$ is determined by
work performed on the system,
\begin{equation}
  \label{eq:73}
  \massdens \dot{\upvarepsilon} =
  \bn \left(\stress^{\ce{T}} \vel\right)
  + \massdens \vel \vb{b}
  + \massdens h
  - \bn \vb{q}
  - \bn \left(\emf\wedge\mmf\right).
\end{equation}
The first two terms stem from the non-kinematic mechanical work
performed on the system, coupling energy and momentum, see also
\cref{eq:7}. The third term describes heat production $h$. As $h$
couples energy and reversible entropy, no constitutive equations for
$h$ is needed. This is consistent because heat production, like
body-forces $\vb{b}$, is an external field. The last two heat-fluxes
are the material heat-flux due to heat-conduction and the
momentum-flux of the electromagnetic field, described by the
Poynting-vector $\emf\wedge \mmf$. Here,
$\emf = \elf + \vel\wedge \magf$ and
$\mmf = \hmagf - \vel\wedge \delf$ are the Galilei-invariant electric
and magnetic fields.\cite{Kinsler2009}

We eliminate the mechanical source-term $\massdens\vel\vb{b}$ by
substitution of \cref{eq:7} into \cref{eq:73}. Using the vector
identity
$\bn ( \stress^{\ce{T}} \vel) = \vel \cdot \bn\stress + \stress :
\bolds{\upkappa}$, we find
\begin{gather}
  \label{eq:9}
  \massdens
  \dot{\upvarepsilon}
  =
  \stress
  :
  \bolds{\upkappa}
  - \bn
  \left(
    \vb{q}
    + \bolds{\mathcal{E}} \wedge \bmc{H}
  \right)
  +
  \massdens
  \left(
    \vel \dot{\vb{g}}
    + h
  \right).
\end{gather}
The Cauchy-stress tensor $\stress$ and the strain-rate-tensor
$\bolds{\upkappa} = ( \ce{grad} \vel + \ce{grad} \ \vel^{\ce{T}})/2$
couple via complete contraction
$\stress : \bolds{\upkappa} = \sum_{ij} \sigmaup_{ij}
\upkappa_{ij}$. The trace of the strain-rate-tensor,
$\ce{tr}\bolds{\upkappa} = \bn\vel$, determines volume-expansion,
whereas its anti-symmetric part describes shear of the
liquid.\cite{poisson_2004}

According to the second axiom of thermodynamics, entropy is not
conserved. We express its evolution with the local Clausius-Duhem
inequality,\cite{truesdell2012rational}
\begin{equation}
  \label{eq:clausius_duhem}
  \massdens\dot{\entro}  \geq   -  \bn \sflux  +  \frac{    \massdens h  }{    \tempr  },
\end{equation}
where $\sflux$ is the non-convective entropy flux density. As measure
for the deviation from thermodynamic equlibrium, \textit{i.e.},
equality in \cref{eq:clausius_duhem}, we define the entropy production
rate $\mathcal{R}$,
\begin{equation}
  \label{eq:12}
  \mathcal{R} = \massdens\tempr\dot{\entro} + T\bn\sflux - \massdens h \geq 0.
\end{equation}
Thus, $\mathcal{R}/\tempr$ is the irreversible part of entropy
production. We want to find expressions for $\mathcal{R}$ and
$\sflux$. To this aim, we replace the energy production $h$ in the
entropy production in \cref{eq:12} with the energy evolution in
\cref{eq:9},
\begin{equation}
\label{eq:5}
\mathcal{R} = \massdens \left(\tempr\dot{\entro}-
  \dot{u}\right)+\tempr\bn\sflux - \bn \vb{q} - \bn
\left(\emf\wedge\mmf\right) \\ 
+ \stress : \bolds{\upkappa}\geq 0.  
\end{equation}
Here, we introduce the internal energy $u$ in the center-of-mass
system, and relate it to the total energy $\upvarepsilon$ with the
differential relation
$\dot{u} = \dot{\upvarepsilon} - \vel\dot{\vb{g}}$.

Our principal modeling-quantity is the free energy density
$\freedens$, which is related to the internal energy $u$ by the
Legendre-transformation $\freedens = u - \tempr\entro$, \textit{i.e.},
\begin{equation}
\label{eq:55}
\freedensdot = \dot u - \tempr \dot\entro - \dot\tempr \entro.  
\end{equation}
The Maxwell equations determine the electromagnetic energy flux
density,
$\bn(\bmc{H}\wedge\bolds{\mathcal{E}}) = \bolds{\mathcal{E}}\Jflux +
\bolds{\mathcal{E}}\dot{\delf}+ \bmc{H}\dot{\magf} +
[(\bolds{\mathcal{E}}\delf + \bmc{H} \magf)\idb -
\bolds{\mathcal{E}}\otimes\delf - \bmc{H}\otimes \magf]:\grada\vel$,
see \cref*{SIsec:generalized-velocity}. Thus, the entropy production
rate becomes
\begin{multline}
\label{eq:60new}
\mathcal{R}
=
- \massdens \freedensdot
- \massdens\entro \dot{\tempr}
+\tempr
\bn\sflux - \bn \vb{q} 
+ \bolds{\mathcal{E}}
\Jflux  
+\bolds{\mathcal{E}}  \dot{\vb{D}}
+ \bmc{H} \dot{\vb{B}} 
\\
+ \bigl[
\stress +
\left(
\emf\delf + \mmf\magf
\right)\idb
- \emf\otimes \delf
- \mmf\otimes \magf
\bigr]
: \grada \vel
\geq 0.
\end{multline}
We continue evaluating the entropy poduction $\mathcal{R}$, and the
entropy flux density $\sflux$ by modeling the free energy density
$\freedens$ in the next section. The final form of $\mathcal{R}$ then
describes dissipative processes and determines our transport
equations.

\subsubsection*{First Law Of Thermodynamics: Energy}
\label{sec:const-equat}

The discussions above are valid for many
materials.\cite{lebon} In this section, we describe
the process of modeling material properties and derive constitutive
equations for the field variables.

The Helmholtz free energy $F$ is the focal point of our material
model. In this work, we study bulk-phenomena and model the
corresponding local free energy density $\freedens$,
\begin{gather}
  \label{eq:66}
  F =   \int_V    \massdens\freedens\ce{d}V.
\end{gather}
Nevertheless, our theory can be extended to non-local phenomena which
are important at electrochemical interfaces. For example, we have
recently shown how to supplement $F$ with an interaction-functional,
incorporating the particle-nature of the
constituents.\cite{C7CP08243F} For a multi-com\-po\-nent, polarizable
and magnetizable liquid,\cite{Henjes1993a} the total differential of
the free energy density reads
\begin{equation}
  \label{eq:67}
  \text{d} (\massdens\freedens)
  =   \emf \cdot \text{d}\delf
    + \mmf \cdot \text{d}\magf
    - \massdens \entro \cdot \text{d}\tempr
    + \sum_{\alpha=1}^\tn   \chempot_\alpha \cdot \text{d}\conc_\alpha.
\end{equation}
Temperature $\tempr$ and concentrations $\conc_\alpha$ are varied
according to standard thermodynamics.\cite{Dreyer2018} The
differential energy
$\emf \cdot \text{d} \delf + \mmf\cdot \text{d}\magf$ with the
independent, Galilei invariant variables $\delf$ and $\magf$ is valid
for linear and non-linear materials.\cite{jackson1999classical}
Previous works, however, have chosen $\emf$\cite{arnulfbs} or
($\vb{P}, \vb{M}$)\cite{Dreyer2018} as independent variables. Since
only unique transformations between the electromagnetic variables
exist, this poses no problem,\cite{hutter2007electromagnetic} but
leads to different expressions for the stress-tensors and the
body-foces employed in the balance-laws.\cite{Reich2018} Varying the
free energy density with respect to the strain-tensor would extend the
model to (visco)-elastic materials, which lies beyond the scope of
this work.

This total differential in \cref{eq:67} consists of products
$X^A\cdot \text{d} \varUpsilon^A$, where $X^A$ describes a particular
reaction of the medium subject to variations of the external quantity
$\varUpsilon^A$. This motivates using
$\bolds{\varUpsilon} = \{\varUpsilon^A\}$ as variable-set for $F$ and
$\freedens$. However, the variational expansion of the free energy
does not capture the complete variable-set, as it does not comprise
viscous, dissipative effects. In particular, viscous liquids cannot
sustain shear stress (unlike elastic media) and depend on the rate of
strain. Due to symmetry-arguments (see section
\cref*{SIsec:orth-transf}), we use the symmetrized, objective strain
rate tensor.\cite{Noll1970a,DeBoor1985,Martins1999} Thus, we model
electrolytes with the objective variable set
$\bolds{\varUpsilon} = \{ \tempr, \conc_\alpha , \delf , \magf,
\bolds{\upkappa} \}$.

The variation of free energy density in \cref{eq:67} contains some
constitutive equations,
\begin{gather}
  \entro = -\frac{\p\freedens}{\p\tempr}, \label{eq:const_entro}\\
  \emf = \massdens \frac{\p\freedens}{\p\delf}, \label{eq:const_delf}\\
  \mmf = \massdens \frac{\p\freedens}{\p\magf}, \label{eq:const_hmagf}\\
  \chempot_\alpha = 
  \frac{\p(\massdens\freedens)}{\p\conc_\alpha}.\label{eq:const_chempot}
\end{gather}
These are supplemented by %the trivial equation
$\p(\massdens\freedens) / \p \upkappa_{ij} {=}
0$. \Cref{eq:const_entro} implies a constitutive equation for the
internal energy density,
$\massdens
u=-\tempr^2\cdot\partial/\partial\tempr(\massdens\freedens/\tempr)$. Constitutive
equations relate conjugate variable-pairs,
$(\entro,\tempr), (\emf,\delf), (\chempot_\alpha, \massdens_\alpha)$,
$(\mmf, \magf)$ and thus complement the universal balance equations
with material-specific properties.

We want to determine $\massdens \freedensdot$ as it enters the entropy
production $\mathcal{R}$ in \cref{eq:60new}. To this aim, the standard
method of Coleman and Noll,\cite{kovetz} assumes
$\freedensdot(\bolds{\varUpsilon})=
\sum_A\p\freedens/\p\varUpsilon^A\cdot \dot{\varUpsilon}^A$. However,
this expansion % is inaccurate and
would result in ambiguous
constitutive equations for the chemical potentials because the
variable-set $\bolds{\varUpsilon}$ is not independent and contains
redundancies (see \cref*{SIsec:supp_const-equat-chem}). Instead, we
determine $\massdens \freedensdot$ from the total differential of the
free energy density (see \cref{eq:67}),
\begin{align}
  \label{eq:74}
  \massdens \freedensdot
  &= \frac{\text{d}(\massdens\freedens)}{\text{d} t}
    -  \dot{\massdens} \freedens  
    =\frac{\text{d}(\massdens\freedens)}{\text{d} t}
    +\massdens\freedens\bn\vel \notag
  \\ 
  &=
   \emf \dot{\delf}
  + \mmf \dot{\magf}
  - \massdens\entro \dot{\tempr}
  + \sum_\alpha \chempot_\alpha \dot{\conc}_\alpha
  + \massdens\freedens \bn\vel,
\end{align}
where we use the mass continuity equation (see
\cref{eq:totmassdens}). Replacing the concentration change
$\dot{\conc}_\alpha$ with the mass balance \cref{eq:concconserv} and
using $\bn\vel=\idb:\bolds{\upkappa}$, we identify the entropy flux
density,\cite{lebon}
\begin{equation}
\sflux=\frac{1}{T} \left( \vb{q}-\chempot_\alpha\Nflux_\alpha\right),
\end{equation}
and the entropy production rate,
\begin{equation}
\label{eq:71}
\mathcal{R} = \bolds{\tauup} : \bolds{\upkappa}
- \sflux  \bn\tempr
- \Jflux \bn\elpot
- \sum_{\alpha=1}^\tn \Nflux_\alpha \bn\chempot_\alpha,
\end{equation}
with the symmetric viscosity tensor
$\bolds{\tauup}(\bolds{\varUpsilon}) = \stress - (\emf\otimes\delf +
\mmf\otimes\magf) + (\emf\delf + \mmf\magf + \sum_{\alpha=1}^\tn
\chempot_\alpha\conc_\alpha - \massdens\freedens) \idb$. To derive
\cref{eq:71}, we use that $\emf\otimes\delf$ and $\mmf\otimes\magf$ is
symmetric, see \cref*{SIsec:symm-visc-tens}.

The entropy production rate $\mathcal{R}$ comprises correlations
between thermodynamic forces and thermodynamic potentials,
$\Nflux_\alpha \bn\chempot_\alpha$, $\Jflux \bn\elpot$ and
$\sflux \bn \tempr$. Positivity of $\mathcal{R}$ constitutes an
important role in our framework, constraining the generalized
fluxes. We proceed by determining the fluxes $\Nflux_\alpha,\sflux$,
and the stress tensor $\stress$ in the next two paragraphs, see
\cref{eq:alternative23,eq:alternative23b,eq:alternative24,eq:mod_stress}.

\subsubsection*{Flux Densities}
\label{sec:onsager-ansatz}

In this section, we determine the flux densities $\Nflux_\alpha, \sflux$
relative to the center-of-mass velocity using the
Onsager-formalism.\cite{grootmazur} This relates the transport equations to
the modeled free energy density $\freedens$.

A dimensional analysis of the Maxwell-equations reveals that magnetic
effects in the bulk-electrolyte can be neglec\-ted under normal
operating conditions of electrochemical systems.\cite{Dreyer2018} We
thus assume from now on the electrostatic limit ($\magf=0$), and
introduce the electric potential $\elpot$ such that
$\emf=\elf=-\bn\elpot$.

Using center-of-mass convection as reference velocity in our transport
theory implies that the trivial flux-constraint 
\begin{equation}
\label{eq:14}
\sum_{\alpha=1}^\tn\mass_\alpha \cdot\Nflux_\alpha=0,
\end{equation}
emerges naturally. Thus, only \tn-1 such fluxes $\Nflux_\alpha$ are
independent, and \cref{eq:14} requires to reduce the number of
species.  To this aim, we define reduced valences and chemical
potentials with respect to the designated species $\alpha=1$,
\begin{gather}
  \label{eq:63}
  \effval_\alpha
  =
  \valence_\alpha
  - \valence_1 \cdot
  \frac{\mass_\alpha}{\mass_1 }, \\
  \effpot_\alpha
  =
  \chempot_\alpha
  - \chempot_1
  \cdot
  \frac{\mass_\alpha}{\mass_1}.
\end{gather}
In the validation chapter, we demonstrate that the predictions of our
theory, if applied correctly, are independent of the choice of the
designated species. This reduction can lead to counter-intuitive
contributions to the electric current,
\begin{equation}
\label{eq:16}
\Jflux = F \sum_{\alpha=2}^\tn \effval_\alpha \cdot \Nflux_\alpha .
\end{equation}
By construction, $\effval_1=0$. Thus the designated species does not
contribute, even if it is charged, $\valence_1\neq 0$. Furthermore, if
the designated species is charged, $\valence_1\neq0$, then even
neutral species carry effective charge and contribute to
$\Jflux$. This is one reason behind recent confusion with respect to
solvent-free electrolytes, \emph{e.g.},
molten-salts,\cite{mamantov2012molten, ratkje1993transference} and
ILs.\cite{C7CP08580J,C8CP02595A} For electrolytes with neutral
solvents, the natural choice for the designated species is the
solvent,\cite{arnulfbs} since then $\effval_\alpha=\valence_\alpha$.

The entropy production rate in the reduced formalism becomes
\begin{align}
  \label{eq:10}
  \mathcal{R} &= \bolds{\tauup}:\bolds{\upkappa}  -
  \sum_{\alpha=2}^\tn \Nflux_\alpha ( F \effval_\alpha \bn \elpot + \bn
  \effpot_\alpha) - \sflux \bn\tempr\\
  \label{eq:quadratic form}
     &=   \bolds{\tauup}:\bolds{\upkappa}  -  \sum_{(\alpha)}
       \bolds{\varPsi}_{(\alpha)}  \cdot  \bolds{X}_{(\alpha)}  \geq
       0. 
\end{align}
Here, we defined the vector of thermodynamic forces
$\bolds{X}_{(\alpha)} = (F \effval_\alpha\bn \elpot +
\bn\effpot_\alpha, \,\bn\tempr)$ and the vector of thermodynamic
fluxes $\bolds{\varPsi}_{(\alpha)} = (\Nflux_\alpha, \, \sflux)$ with
components $\varPsi^A_{(\alpha)}$. The subscripts in brackets denote
that only the ionic contributions are species-related.

We make use of this expansion and transfer the products to quadratic
terms via a bilinear Onsager-matrix. Thus, this Onsager formalism (i)
closes the system of equations, (ii) establishes positivity of
$\mathcal{R}$, and (iii) evaluates the flux-force couplings. The
phenomenological relations linearly couple thermodynamic fluxes and
forces with the Onsager matrix $\mathcal{L}_{(\alpha)(\beta)}$,
\begin{gather}
  \label{eq:27}
  \bolds{\varPsi}^A_{(\alpha)}
  = - \sum_{\beta=2}^\tn
  \mathcal{L}_{(\alpha)(\beta)} \bolds{X}_{(\beta)},
  \\ 
  \label{eq:26}
  \left(
    \begin{matrix}
      \Nflux_2 \\ \vdots \\ \Nflux_\tn \\ \sflux
    \end{matrix}
  \right)
  =
  - \left(
    \begin{matrix}
      \mathcal{L}_{22} & \ldots & \mathcal{L}_{2\tn} &
      \mathcal{L}_{2\tempr}
      \\
      \vdots & \ddots & \vdots & \vdots
      \\
      \mathcal{L}_{2\tn} & \ldots & \mathcal{L}_{\tn\tn} &
      \mathcal{L}_{\tn\tempr}
      \\
      \mathcal{L}_{2\tempr} & \ldots & \mathcal{L}_{\tn\tempr} &
      \mathcal{L}_{\tempr\tempr}
    \end{matrix}
  \right) 
  \cdot
  \left(
    \begin{matrix}
      F \effval_2\bn \elpot+\bn\effpot_2
      \\ \vdots
      \\ F \effval_\tn\bn
      \elpot+\bn \effpot_\tn
      \\\bn\tempr
    \end{matrix}
  \right).
\end{gather}
The Onsager matrix $\mathcal{L}_{(\alpha)(\beta)}$ is symmetric and
positive semi-definite.\cite{dreyer} This guarantees that the
corresponding term in the entropy production rate $\mathcal{R}$ is
always positive (see \cref{eq:quadratic form}). Thus, electrolyte
transport is a dissipative process, which wants to equilibrate the
system. In dilute solutions, the Onsager matrix is diagonal and
describes intra-species correlations. In concentrated electrolytes,
off-diagonal Onsager coefficients comprise inter-species correlations,
\emph{e.g.}, electro-osmotic drag through a membrane \cite{Monroe2017}
or asymmetric transference numbers.\cite{wohde2016li+} In total, we
define $\tn(\tn+1)/2$ independent Onsager coefficients, \textit{i.e.},
transport parameters.

The thermodynamic fluxes are driven by electric potential gradients,
concentration gradients, and temperature gradients, \textit{i.e.},
migration, diffusion, and thermo-electric effects. In the following,
we relate the abstract Onsager coefficients to these physico-chemical
effects. The electric current density $\Jflux$ (see \cref{eq:16}) is
expressed as
\begin{equation}
\label{eq:alternative20b}
  \Jflux = -\elcond\bn\elpot -\elcond \sum_{\beta=2}^\tn
  \frac{\trans_\beta}{F\effval_\beta} \bn \effpot_\beta 
  -\elcond \upbeta\bn\tempr.  
\end{equation}
The flux densities $\Nflux_\alpha$ and the entropy flux density
$\sflux$ are often formulated in terms of $\Jflux$ instead of
$\elpot$. Therefore, we transform \cref{eq:26} by substituting
$\bn\elpot$ for $\Jflux$ with \cref{eq:alternative20b} and introducing
additional, physically motivated parameters,
\begin{gather}
  \label{eq:iofl}
  %% Two-column style
% \begin{multlined}
%   \Nflux_\alpha =
%   \frac{ \trans_\alpha \Jflux }{ F \effval_\alpha }
%   - \sum_{\beta=2}^\tn \left( \mathcal{L}_{\alpha\beta} -\frac{\elcond
%       \trans_\alpha\trans_\beta}{F^2 \effval_\alpha\effval_\beta}
%   \right) \bn\effpot_\beta
%   \\
%     - \left(
%     \mathcal{L}_{\alpha\tempr}
%     - \frac{ \upbeta \elcond \trans_\alpha }{ F\effval_\alpha }
%   \right) \bn\tempr,
% \end{multlined}
%% Single-column style
  \Nflux_\alpha =
  \frac{ \trans_\alpha \Jflux }{ F \effval_\alpha }
  - \sum_{\beta=2}^\tn \left( \mathcal{L}_{\alpha\beta} -\frac{\elcond
      \trans_\alpha\trans_\beta}{F^2 \effval_\alpha\effval_\beta}
  \right) \bn\effpot_\beta
    - \left(
    \mathcal{L}_{\alpha\tempr}
    - \frac{ \upbeta \elcond \trans_\alpha }{ F\effval_\alpha }
  \right) \bn\tempr,
\\
\label{eq:ioflb}
\sflux = \upbeta\Jflux-
\left( \frac{ \heatcond }{ \tempr } - \upbeta^2 \elcond \right)
\bn\tempr - \sum_{\beta=2}^\tn \left( \mathcal{L}_{\beta\tempr} -
  \upbeta \frac{ \elcond\trans_\beta }{ F \effval_\beta } \right)
\bn\effpot_\beta.
\end{gather}
Transport coefficients introduced in
\cref{eq:alternative20b,eq:iofl,,eq:ioflb} are electric conductivity
$\elcond$, Seebeck coefficient $\upbeta$, and thermal conductivity
$\heatcond$, \cite{Newman2004}
\begin{gather}
  \label{eq:def_elcond}
  \elcond= F^2 \sum_{\alpha,\beta=2}^\tn   \mathcal{L}_{\alpha\beta}
  \effval_\alpha\effval_\beta   ,  \\ 
  \label{eq:def_seebeck}
  \upbeta = \frac{F  }{    \elcond  }  \sum_{\alpha=2}^\tn
  \mathcal{L}_{\alpha    \tempr} \effval_\alpha,  \\ 
  \label{eq:def_heatcond}
  \heatcond = \tempr \mathcal{L}_{\tempr\tempr}.
\end{gather}
By construction, $\elcond$ and $\heatcond$ are positive because the
Onsager matrix is positive semi-definite. $\tn$-1 transference numbers
$\trans_\alpha$ relate particle fluxes and electric current in
\cref{eq:iofl},
\begin{equation}
\label{eq:def_transferencenumber}
\trans_\alpha = \frac{ 	F^2 \effval_\alpha }{ 	\elcond }
\sum_{\beta=2}^\tn \mathcal{L}_{\alpha\beta} \effval_\beta.  
\end{equation}
These \tn-1 parameters are subject to the normalization condition,
$\sum_{\alpha=2}^\tn \trans_\alpha = 1$, such that only \tn-2
transference numbers are independent. Thus, for binary electrolytes
all transference numbers are fixed, $\trans_2^{\tn=2}=1$, and
$\Nflux_1$ relates to $\Jflux$ by the specific mass-ratio alone, see
\cref*{SIsec:comm-transf-numb}. Apparently, all $\trans_\alpha$ are
positive if the Onsager matrix is diagonal (as for dilute solutions).

We proceed by defining the \tn(\tn+1)/2 coefficients of the symmetric
diffusion matrix $\diffusion$,
\begin{gather}
\diffusioncomp_{\alpha\beta} = \mathcal{L}_{\alpha\beta} - \frac{
  \elcond\trans_\alpha\trans_\beta 
}{	F^2 \effval_\alpha	\effval_\beta} 
\label{eq:D_ab}, \nonumber \\
\diffusioncomp_{\alpha\tempr} = \mathcal{L}_{\alpha\tempr} - \frac{
  \upbeta\elcond\trans_\alpha}{	F \effval_\alpha}, 
\label{eq:D_aT}\\
\diffusioncomp_{\tempr\tempr}=\frac{	\heatcond}{	\tempr}-
\upbeta^2 \elcond 
\label{eq:D_TT}.\nonumber
\end{gather}
Since $\diffusioncomp_{\tempr\tempr}$ is determined by
$\heatcond,\upbeta$ and $\elcond$,
\cref{eq:def_elcond,eq:def_seebeck,eq:def_heatcond,eq:def_transferencenumber,eq:D_aT}
yield \tn(\tn+3)/2 transport coefficients. Thus, the number of
transport parameters defined so far exceeds the number of independent
Onsager coefficients $\tn(\tn+1)/2$. However, further $\tn$
constraints follow from the relation of $\Jflux$ and $\Nflux_\alpha$
in \cref{eq:16},
\begin{gather}
  \label{eq:24}
  \sum_{\beta=2}^\tn  \diffusioncomp_{\alpha\beta}  \effval_\beta  =  0,
  \qquad
  \sum_{\beta = 2}^\tn  \diffusioncomp_{\tempr\beta}  \effval_\beta  =  0.
\end{gather}
Thus, only \tn(\tn-1)/2 diffusion coefficients are independent, where
$ \diffusioncomp_{2\alpha} {=} - \sum_{\beta=3}^\tn
\diffusioncomp_{\alpha\beta} \effval_\beta / \effval_2 $ and
$ \diffusioncomp_{2\tempr} {=} -\sum_{\beta=3}^\tn
\diffusioncomp_{\tempr\beta} \effval_\beta / \effval_2 $. In
particular
$\diffusioncomp_{22} {=} \sum_{\beta,\gamma=3}^\tn
\diffusioncomp_{\beta\gamma} \effval_\beta \effval_\gamma
/(\effval_2)^2 $.  Altogether, \tn(\tn-1)/2 independent diffusion
coefficients, \tn-2 independent transference numbers, the electric
conductivity, and the Seebeck coefficient constitute the complete set
of physically motivated free parameters.

We designate a second species, $\alpha=2$, and define the set of \tn-2
reduced chemical potentials
\begin{equation}
  \label{eq:62}
  \effeffpot_\beta
  = \effpot_\beta -   \frac{ \effval_\beta }{ \effval_2 } \cdot \effpot_2.
\end{equation}
Thus \tn-1 independent fluxes ($\Nflux_3,\ldots,\Nflux_\tn,\sflux$) exist, 
\begin{gather}
    \label{eq:alternative23}
  \Nflux_\alpha  =  \frac{    \trans_\alpha  }{    F\effval_\alpha
  }  \Jflux - \diffusioncomp_{\alpha\tempr} \bn\tempr  -
  \sum_{\beta=3}^\tn  \diffusioncomp_{\alpha\beta}\bn\effeffpot_\beta,
  \quad \alpha\geq 3\\
  \label{eq:alternative23b}
  \sflux  =  \upbeta  \Jflux  -\diffusioncomp_{\tempr\tempr}  \bn\tempr  -  \sum_{\beta=3}^\tn
  \diffusioncomp_{\beta\tempr}  \bn \effeffpot_\beta  . 
\end{gather}
$\Nflux_2$ is determined by the electric flux with \cref{eq:16} and
$\Nflux_1$ is determined by mass conservation in \cref{eq:14}. Our
electrolyte potential $\elpot$ is the Maxwell potential that appears
in the Poisson equation. Typically,\cite{arnulfbs,Newman2004}
concentrated solution theory is expressed using the alternative
potential
\begin{equation}
  \label{eq:electrochempoteff}
  \varphi ( \elpot, \chempot_{1,2}, \mass_{1,2},\valence_{1,2} )
  = \elpot + \effpot_2 / F \effval_2. 
\end{equation}
If the first designated species is the neutral solvent, 
$\varphi = \elpot + \effpot_2 /F\valence_2$ corresponds to the
electro-chemical potential of the second designated species. With the
electrolyte potential $\varphi$, we can express the current density analogous to \cref{eq:alternative23}
and \cref{eq:alternative23b},
\begin{equation}
  \label{eq:alternative24}
  \Jflux   =  - \elcond\bn   \varphi  - \upbeta\elcond \bn\tempr  -  \frac{  \elcond  }{  F  }  \sum_{\beta=3}^\tn
  \frac{  \trans_\beta  }{  \effval_\beta  }  \bn \effeffpot_\beta.
\end{equation}
The entropy production rate $\mathcal{R}$ in the double reduced formalism is
\begin{align}
  \label{eq:R_final}
  \mathcal{R}
  &=  
    \bolds{\tauup}:\bolds{\upkappa}
    - \Jflux\bn\varphi
    - \sum_{\alpha=3}^\tn \Nflux_\alpha  \bn \effeffpot_\alpha- \sflux
    \bn\tempr
  \\
  &=
    \begin{multlined}[t]
      \bolds{\tauup}:\bolds{\upkappa} + \Jflux^2/\elcond +
      \\
      + (\bn\effeffpot_3,\ldots\bn\effeffpot_\tn,\bn\tempr)
      \cdot\mathcal{D}^\text{red}_{(\alpha),(\beta)}\cdot
      (\bn\effeffpot_3,\ldots\bn\effeffpot_\tn,\bn\tempr)^\text{T}
      \nonumber 
    \end{multlined}
  \\
  &\geq 0,
\nonumber
\end{align}
where $\mathcal{D}^\text{red}$ is the diffusion matrix $\mathcal{D}$
without the two designated species (see
\cref{eq:D_aT}). To summarize, the entropy production
comprises three contributions. The first term describes mechanical
dissipation. In the next paragraph, we determine $\bolds{\tauup}$ in such a way
that $\bolds{\tauup}:\bolds{\upkappa} \geq 0$ is guaranteed. The
second term describes Joule-heating due to migration. Since the
electric conductivity is non-negative, $\Jflux^2/\elcond\geq 0$. The
last term describes entropy production due to diffusion and heat
conduction. Thermodynamic consistency reduces to non-negativity of the
diffusion matrix, $\mathcal{D}^\text{red}\geq 0$.

\subsubsection*{Stress Tensor}
\label{sec:stress}
In contrast to viscous-elastic models, this description does not
describe the stress tensor by a constitutive equation in the form of a
derivative of the free energy density.\cite{Huetter2016,lebon}
Instead, we determine $\stress$ implicitely via the viscosity
tensor. To find $\bolds{\tauup}(\bolds{\upkappa})$, we apply the
Onsager formalism as in the previous subsection, and assume that
$\bolds{\tauup}$ is linear in the strain-rate tensor. Symmetry demands
that $\bolds{\tauup}$ is an objective tensor, thus the representation
theorems for isotropic tensors determine it uniquely up to two scalar
transport parameters of viscosity,\cite{Coleman1963,batchelor_2000}
\begin{equation}
\label{eq:72}
\bolds{\tauup} = \lambda(\tempr, \massdens_\alpha) (\bn \vel) \idb +   2\eta(\tempr , \massdens_\alpha )
\bolds{\upkappa}_\text{tf},
\end{equation}
where $\bolds{\upkappa}_{\ce{tf}}$ is the symmetric trace-free part of
$\bolds{\upkappa}$, $\lambda$ is the bulk-viscosity and $\eta$ is the
shear-viscosity. Since
\begin{equation}
  \label{eq:const_eq_viscosity}
  \bolds{\tauup}:\bolds{\upkappa} =
  \lambda(\bn\vel)^2 + 2\eta\bolds{\upkappa}_{\ce{tf}}:\bolds{\upkappa}_{\ce{tf}}\ge 0,
\end{equation}
the entropy production rate $\mathcal{R}$ is non-negative (see
\cref{eq:71}), if the viscosities obey $\eta \geq 0$ and
$\lambda \geq 0$. To summarize, we find as constitutive equation for
the symmetric stress tensor,\cite{binnemans2005ionic}
\begin{equation}
  \label{eq:mod_stress}
  \stress
  =
    \elf   \otimes  \delf
  +   \left(
    \massdens \freedens      - \sum_{\alpha=1}^\tn
    \conc_\alpha    \chempot_\alpha     - \elf\delf  + \lambda \bn
    \vel  \right)  \idb 
  + 2\eta \bolds{\upkappa}_\text{tf}.
\end{equation}

The pressure $p$ measured in experiments is derived from the total
stress tensor,\cite{becker1933theorie,Dreyer2018}
\begin{equation}
  \label{eq:mean_normal_stress}
  p
  =
  - \frac{1}{3} \ce{tr}(\stress)
  =
  p^{\ce{td}}
  - \lambda\cdot\bn\vel. 
\end{equation}
Here,
$p^{\ce{td}} = \sum_{\alpha=1}^\tn \conc_\alpha\chempot_\alpha -
\massdens \freedens + 2\elf\delf/3$ is the thermodynamic pressure
comprising the standard Gibbs-Duhem contribution, and a
Maxwell-contribution. The further contribution, $\lambda\bn\vel$,
describes dissipation due to friction. Since $\lambda\geq 0$, friction
can lead to positive and negative pressures, depending on $\bn\vel$.

The apparent electromagnetic contribution to the total stress
$\stress$ in \cref{eq:mod_stress} differs from the Maxwell
contribution $\bolds{\Sigma} = - \elf\delf/2 + \elf\otimes\delf$. Note
that we recover the correct standard form once we model and specify
$\freedens$ (see \cref{eq:stress}).

\subsection*{Model for Correlated Liquid Electrolytes}
\label{sec:free-enrgy-modelling}
The transport equations of any electrolyte-model depend on the
specific form of the free energy density $\freedens$. In our approach,
$\freedens$ is the focal point of modeling. Once $\freedens$ is
specified, all equations follow from plain mathematics. In this
section we introduce our model $\freedens$,
derive the convection velocity from the volumetric electrolyte
equation of state, and discuss the
dynamic transport equations.

\subsubsection*{Free Energy Density}
\label{sec:free-energy}
As discussed in the previous section, we assume that the free energy
density $\freedens$ depends on temperature $\tempr$, concentrations
$\conc_\alpha$, and dielectric displacement $\delf$ (see \cref{eq:67}
and the discussion thereafter).\cite{Dreyer2018,arnulfbs} Our model
for liquid electrolytes is generated by
\begin{multline}
  \label{eq:4}
%%  Two column style
  \massdens\freedens
  =  \frac{\delf^2}{2\diel(1+\suszept)} 
  +  \frac{\bulkmodalt}{2}  \left(    1 - \sum_{\alpha=1}^\tn    \pmv^0_\alpha
    \conc_\alpha  \right)^2 
  +  R\tempr  \sum_{\alpha=1}^\tn  \conc_\alpha
  \ln\left( \frac{ \conc_\alpha  }{    \conc  }  \right)
  \\
  +   \massdens\freedens^{\ce{int}}.
\end{multline}
% \begin{equation}
%   %% Single column style
%   \label{eq:4}
%     \massdens\freedens
%   =  \frac{\delf^2}{2\diel(1+\suszept)} 
%   +  \frac{\bulkmodalt}{2}  \left(    1 - \sum_{\alpha=1}^\tn    \pmv^0_\alpha
%     \conc_\alpha  \right)^2 
%   +  R\tempr  \sum_{\alpha=1}^\tn  \conc_\alpha
%   \ln\left( \frac{ \conc_\alpha  }{    \conc  }  \right)
%   +   \massdens\freedens^{\ce{int}}.
% \end{equation}
The first term comprises the electrostatic energy-density of
polarizable media.\cite{jackson1999classical} The eletric field
follows from \cref{eq:const_delf},
$\elf = \delf / \diel (1+\suszept)$, which describes a linear
dielectric medium. We neglect the dependence of susceptibility
$\suszept$ on ion composition such that the chemical potentials do not
depend on dielectric displacement (see section
\cref*{SIsec:supp_heat-equation}).

The second term expresses volumetric contributions with partial molar
volumes $\pmv^0_\alpha$ in a stable reference state, which couple to
surface forces acting upon the system. These surface forces are
usually expressed by pressure (see \cref{eq:stress}). Here,
$\bulkmodalt$ is the bulk-modulus which acts as a Lagrange-mutliplier
in the case of incompressible electrolytes (see
\cref{eq:pmv_zero_const}). We expand the energy of deformations around
a stable reference state in order to easily transfer to
incompresssible media, see \cref{eq:pmv_zero_const}. These volumetric
energy penalties encode volume conservation (see \cref{eq:38}), and
account for mean steric effects,\cite{Fedorov2014} which are crucial
for continuum theories of concentrated electrolytes \cite{Freise1952}
and ILs.\cite{doi:10.1021/acs.jpclett.7b03048} Mean steric effects
create a threshold for local ion-concentration and prevent Coulombic
collapse of the system.\cite{hansen2006theory} This leads to crowding
of ILs near electrified interfaces.\cite{C7CP08243F} We motivate this
expression for the elastic energy in the supplementary
(\cref*{SIsec:supp_modell-free-energy-1}).

The third term is the entropy of mixture for non-interacting systems. We model
this term in analogy with ideal gases and neglect contributions from
inter-molecular interactions, \emph{e.g.}, solvation effects.

% Instead of using a modified statistics,\cite{Dreyer2014} 
We phenomenologically account for non-ideal interaction contributions
in $\massdens\freedens^{\ce{int}}$. The activity-coefficients
$ \textsf{\textsl{f}}_\alpha$ measure the deviation from ideal/dilute
electrolytes ($ \textsf{\textsl{f}}_\alpha\conc{=}1$) via
$\p(\massdens\freedens^{\ce{int}})/\p\conc_\alpha=R\tempr\ln(
\textsf{\textsl{f}}_\alpha\conc)$. The interaction term
$\massdens\freedens^{\ce{int}}$ can also describe the heat capacity of
the system,
$\textsf{\textsl{{C}}} = - \tempr \cdot \p^2 ( \massdens
\freedens^{\ce{int}}) / \p \tempr^2$. Since thermal aspects are not
our main focus, we refer to the supplementary for a thermal
contribution to our free energy model $\massdens\freedens$
(\cref*{SIsec:thermal-model-free}).

In the following, we calulate the chemical potentials
$\chempot_\alpha$ and the stress tensor from the model free energy
density in \cref{eq:4}. The chemical potentials follow by evaluation
of the constitutive \cref{eq:const_chempot},
\begin{equation}
\label{eq:chempot}
\chempot_\alpha 
= R \tempr \ln( \textsf{\textsl{f}}_\alpha\conc_\alpha) 
- \bulkmodalt
\pmv^0_\alpha \left( 1- \sum_{\beta=1}^\tn \pmv_\beta^0 \conc_\beta
\right).  
\end{equation}
The first term comprises entropy of mixture and inter-mo\-le\-cu\-lar
interaction energies. The stress tensor is determined by
\cref{eq:mod_stress},
\begin{multline}
  %% Two column style
  \label{eq:stress}
  \stress
  =
  \elf\otimes\delf 
  - \frac{\elf\delf}{2} \idb 
  + \frac{\bulkmodalt}{2} \idb
  - \frac{\bulkmodalt}{2}
  \left(
    \sum_{\alpha=1}^\tn\conc_\alpha\pmv_\alpha^0
  \right)^2\idb
  + \lambda \cdot\bn \vel \idb
  \\
  + 2\eta \bolds{\upkappa}_\text{tf}
  - \massdens\sum_{\alpha=1}^\tn\conc_\alpha
  \frac{\partial \freedens^{\text{int}}}{\partial \conc_\alpha} \idb.
\end{multline}
% \begin{equation}
%   %% Single column style
%   \label{eq:stress}
%   \stress
%   =
%   \elf\otimes\delf 
%   - \frac{\elf\delf}{2} \idb 
%   + \frac{\bulkmodalt}{2} \idb
%   - \frac{\bulkmodalt}{2}
%   \left(
%     \sum_{\alpha=1}^\tn\conc_\alpha\pmv_\alpha^0
%   \right)^2\idb
%   + \lambda \cdot\bn \vel \idb
%   + 2\eta \bolds{\upkappa}_\text{tf}
%   - \massdens\sum_{\alpha=1}^\tn\conc_\alpha
%   \frac{\partial\freedens^{\text{int}}}{\partial \conc_\alpha} \idb.
% \end{equation}
Thus, the electrostatic contribution is the expected electrostatic
Maxwell-stress tensor
$\boldsymbol{\Sigma}$.\cite{jackson1999classical} The elastic stress
due to compressibility and the stress due to intra-molecular
interaction is also diagonal, \textit{i.e.}, isotropic. Shear stress due to
viscosity contains the non-isotropic contribution
$2\eta\bolds{\upkappa}_\text{tf}$.

\subsubsection*{Volume Constraint and Convection}
\label{sec:constr-incompr-mass}
In this section, we derive the equation of state for liquids from our
model free energy (see \cref{eq:4}), and discuss incompressibility of
electrolytes. In addition, we derive a multi-component
incompressibility constraint, and a transport equation for the
center-of-mass convection velocity $\vel$.  

In a homogeneous system we neglect viscosity. In this case
$p=p^{\ce{td}}$ becomes the thermodynamic pressure as discussed for
(\ref{eq:mean_normal_stress}). \Cref{eq:stress} determines,
\begin{equation}
  \label{eq:pressure}
  p = p^{\ce{td}}=-\text{tr}(\stress|_{\bn\vel=0})/3=f(\conc_\alpha,\tempr,\delf),
\end{equation}
as a function of $\conc_\alpha$, $\tempr$, and $\delf$. This
expression implicitly determines volume
$V(\mathcal{N}_\alpha,p,\tempr,\delf)$ as function of particle numbers
$\mathcal{N}_\alpha$, pressure $p$, temperature $\tempr$, and
dielectric displacement $\delf$. We make use of this description, and
define partial molar volumes $\pmv_\alpha$ as derivative of $V$ with
respect to $\mathcal{N}_\alpha$. Implicit differentiation of the
function $f(\conc_\alpha,\tempr,\delf)$ then determines the partial
molar volumes (see section \cref*{SIsec:deriv-part-molar}),
\begin{equation}
\label{eq:pmv}
\pmv_\alpha
=
\frac{\partial V}{\partial \mathcal{N}_\alpha}\big|_{\tempr,p,\delf}
=
\frac{\partial f/\partial \conc_\alpha}{\sum_{\beta=1}^\tn \partial f/\partial
  \conc_\beta \cdot \conc_\beta}\big|_{\tempr,p,\delf} .
\end{equation}
Thus, the partial molar volumes of the electrolyte follow from the
stress tensor.

As $V(\mathcal{N}_\alpha,p,\tempr,\delf)$ is a homogeneous function of
first order in $\mathcal{N}_\beta$, Euler's homogeneous function
theorem yields a compact form of the electrolyte equation of state of
liquid
electrolytes,\cite{Freise1952,Horstmann2012a,Dreyer2015,single2017revealing,doi:10.1021/acs.jpclett.7b03048}
\begin{equation}
\label{eq:38}
V=\sum_{\alpha=1}^\tn\pmv_\alpha\mathcal{N}_\alpha ~~\text{   or   }~~
1=\sum_{\alpha=1}^\tn\pmv_\alpha \conc_\alpha, 
\end{equation}  
which takes the form of a volume constraint.

From \cref{eq:pmv}, we calculate the partial molar volumes for our
model. Liquid electrolytes are hardly compressible, \textit{i.e.},
$\bulkmodalt\gg p$. Repeated application of this approximation yields
\begin{align}
  \pmv_\alpha
  &\approx
    \frac{\pmv^0_\alpha}{\sum_{\beta=1}^{\tn}\pmv_\beta^0 \conc_\beta}
  \\
  &\approx\pmv^0_\alpha
    - \frac{\pmv^0_\alpha}{\bulkmodalt}
    \left(
    p
    - \frac{\elf\delf}{6}
    - \massdens \sum_{\alpha=1}^\tn\conc_\alpha
    \frac{\partial\freedens^{\text{int}}
    }{
    \partial\conc_\alpha
    }
    \right).    \label{eq:pmv_zero_const}
\end{align}
Here, we used \cref{eq:pressure} and \cref{eq:stress} to replace
$\sum_{\beta=1}^{\tn}\pmv_\beta^0 \conc_\beta$. Thus, $\bulkmodalt$ is
indeed the bulk-modulus, and $\pmv^0_\alpha$ is the partial molar
volume at standard pressure $p=0$. For incompressible electrolytes
$\bulkmodalt\rightarrow\infty$, the partial molar volumes
$\pmv_\alpha=\pmv^0_\alpha$ do not depend on pressure, and
\cref{eq:38} constitutes an incompressibility
constraint.\cite{C7CP08243F} In this case, pressure cannot be
determined by \cref{eq:pressure}. Nevertheless, the volumetric
contributions in the chemical potentials have to be determined (see
\cref{eq:chempot}). We show in \cref{eq:13}, how to solve this
challenge with the volumetric contributions in the stress-tensor and
momentum conservation.

The volume constraint in \cref{eq:38} remains fulfilled under
multi-component transport as the center-of-mass motion balances the
volumes. Thus, we can derive an equation for the velocity $\vel$ from
the volume constraint. To this aim, let us first proof a
Lemma. Partial molar volumes $\pmv_\alpha$ defined in \cref{eq:pmv}
obey the symmetry-property
$\p\pmv_\alpha/\p \mathcal{N}_\beta =
\p\pmv_\beta/\p\mathcal{N}_\alpha$. Applying Euler's homogeneous
function theorem to $\pmv_\alpha$ ensures that
$ \sum_{\beta=1}^\tn \conc_\beta \p\pmv_\alpha / \p\mathcal{N}_\beta =
0$ holds for all species. These relations proof the following Lemma,
\begin{equation}
\label{eq:41}
\sum_{\alpha=1}^\tn
\conc_\alpha\dot{\pmv}_\alpha = \sum_{\alpha=1}^\tn \frac{ 	\p
  \left( 	\conc_\alpha \pmv_\alpha 	\right) }{
  \p\tempr }\cdot \dot{\tempr}+\sum_{\alpha=1}^\tn\sum_{\beta=1}^\tn 
\conc_\alpha\frac{	\p\pmv_\alpha}{
  \p\mathcal{N}_\beta}\cdot\dot{\mathcal{N}}_\beta= 0, 
\end{equation}
where we used the independence of the primary variables
$\conc_\alpha,\tempr$ and the volume constraint.  

Applying the total time derivative to the volume constraint in
\cref{eq:38}, we conclude from our Lemma that
$\sum_{\alpha=1}^\tn \dot{\conc}_\alpha \pmv_\alpha =0$. In
combination with mass conservation \cref{eq:concconserv} for the
concentrations $\conc_\alpha$, we finally derive the equation for the
center-of-mass velocity $\vel$,
\cite{Stamm2017,Single2019,Schmitt2019}
\begin{equation}
  \label{eq:convectionvel}
  \bn \vel  =  -  \sum_{\alpha=1}^\tn  \pmv_\alpha  \bn  \Nflux_\alpha
  +  \sum_{\alpha=1}^\tn   \pmv_\alpha   r_\alpha.  
\end{equation}
This equation is generally true for compressible as well as
incompressible electrolytes and for constant as well as non-constant
partial molar volumes. 

The left-hand side of \cref{eq:convectionvel} expresses local,
isotropic volume-expansion, $\bn\vel=\ce{tr}( \bolds{\upkappa})$. Thus,
the first term on the right measures volume-expansion caused by
transport. In absence of the second term, and for constant partial
molar volumes, the total flux of volume-fractions
$\conc_\alpha\pmv_\alpha\vel_\alpha$ measured in the fixed frame is
conserved,
$\bn \sum_{\alpha=1}^\tn \conc_\alpha \pmv_\alpha \vel_\alpha=0$. The
second term in \cref{eq:convectionvel} measures volume-production by
chemical reactions, and is an important source for convection in
multicomponent systems.

In the case of a single-component liquid, $\bn\vel=0$ according to
\cref{eq:convectionvel}, since by construction the only flux density
relative to the center of mass motion vanishes, $\Nflux_1=0$. This is
a standard-condition for incompressibility of non-reactive
media. However, it is often used for complex electrolyte-mixtures
which can be a bad approximation.

From now on we consider electrolytes in the incompressible limit
$\bulkmodalt\to\infty$, such that $\pmv_\alpha^0=\pmv_\alpha$ becomes
the general partial molar volume (defined for any state).

As we showed above, conservation of mass and charge reduce the number
of independent species by two. We assign the designated two physical
species to $\alpha=1$, $\alpha=2$. The volume constraint \cref{eq:38}
allows to determine the corresponding concentrations,
\begin{align}
  \label{eq:8}
  \begin{split}
    \conc_1\left(\conc_3,\ldots,\conc_\tn,\chargedens\right)
    &=  \frac{
      1
      - \pmv_2\conc_2
      -\sum_{\alpha=3}^\tn
      \pmv_\alpha
      \conc_\alpha
    }{
      \pmv_1
    },
    \\
    \conc_2\left(\conc_3,\ldots,\conc_\tn,\chargedens\right)
    &=
    \frac{
      \valence_1
      - \pmv_1
      \chargedens/ F 
    }{
      \pmv_2\valence_1
      - \pmv_1\valence_2 
    }
    + \sum_{\alpha=3}^\tn
    \conc_\alpha
    \frac{
      \pmv_\alpha\valence_1
      - \pmv_1 \valence_\alpha  
    }{
      \pmv_1 \valence_2 - \pmv_2\valence_1
    },
  \end{split}
\end{align}
with the charge density $\chargedens$. Furthermore, we transform
\cref{eq:convectionvel} into the reduced description,
\begin{equation}
  \label{eq:2}
  \bn\vel   =   - \frac{\tilde{\pmv}_2}{F \effval_2}\bn \Jflux  -
  \sum_{\alpha=3}^\tn \tilde{\tilde{\pmv}}_\alpha  \bn \Nflux_\alpha, 
\end{equation}
where $\tilde{\pmv}_\alpha = \pmv_\alpha-M_\alpha/M_1\cdot \pmv_1$ and
$\tilde{\tilde{\pmv}}_\alpha=\tilde{\pmv}_\alpha-
\effval_\alpha/\effval_2\cdot \tilde{\pmv}_2$ define the set of independent
partial molar volumes. In particular, \cref{eq:2} implies constant
convection-profiles for electroneutral, binary
electrolytes. Conversely, this also shows that pure ILs cannot sustain
concentration gradients.\cite{mamantov2012molten}

\subsubsection*{Equations of Motion}
\label{sec:equations-motion}
The force law follows from momentum-balance in \cref{eq:7} and takes
the form of a generalized Navier-Stokes equation. However, we assume
highly effective momentum-dissipation in such viscous media. We
neglect external body forces,\cite{dreyer2}, and assume mechanical
equilibrium,\cite{C7CP08243F}
\begin{multline}
    %% Two column style
  \label{eq:navier_stokes}
  0 =\bn \stress
  =  -  \bulkmodalt  \bn \sum_{\alpha=1}^\tn   \conc_\alpha
  \pmv_\alpha
  + \chargedens \elf
  + \eta \bn^2 \vel  + \left(\lambda    + \eta/3 \right) \bn  \left(
    \bn \vel  \right) \\
  - \sum_{\alpha=1}^\tn \conc_\alpha \bn
  \frac{\partial(\massdens\freedens^{\text{int}})}{\partial\conc_\alpha},
\end{multline}
% \begin{equation}
%   %% Single column style
%   \label{eq:navier_stokes}
%   0 =\bn \stress
%   =  -  \bulkmodalt  \bn \sum_{\alpha=1}^\tn   \conc_\alpha
%   \pmv_\alpha
%   + \chargedens \elf
%   + \eta \bn^2 \vel  + \left(\lambda    + \eta/3 \right) \bn  \left(
%     \bn \vel  \right) 
%   - \sum_{\alpha=1}^\tn \conc_\alpha \bn
%   \frac{\partial(\massdens\freedens^{\text{int}})}{\partial\conc_\alpha},
% \end{equation}
Above we assumed constant viscosity-parameters, constant bulk-modulus,
and isothermal conditions. Although the force law is not used in most
of the battery literature,\cite{arnulfbs} it plays a fundamental role
for highly concentrated electrolytes,\cite{dreyer2} and ionic
liquids,\cite{C7CP08243F} where Coulomb interactions compete with
saturation effects.

Stress couples to transport via the volumetric term in the chemical
potentials (see \cref{eq:chempot}). Via this mechanism, we ensure that
the complete, coupled set of mechanical and electrostatic stresses is
comprised in our theory. By substitution, dissipative electric,
viscous, and interaction forces contribute to the chemical potentials,
% \begin{multline}
%   %% Two column style
%   \label{eq:13}
%   \bn\chempot_\alpha 
%   =
%   - \chargedens \pmv_\alpha \bn\elpot
%   + R \tempr  \sum_{\beta}
%   \textsf{\textsl{TDF}}_{\alpha\beta}
%   \frac{
%     \bn \conc_\beta
%   }{
%     \conc_\beta
%   } 
%     \\ 
%     + \pmv_\alpha
%   \left(
%     \left( \lambda
%       + \eta/3
%     \right)
%     \bn
%     \left( \bn\vel
%     \right)
%     + \eta \bn^2 \vel
%   \right),
% \end{multline}
% \noindent \where
% \begin{equation}
%   \label{eq:tdf}
%   \cos  \textsf{\textsl{TDF}}_{\alpha\beta} =
%   \sum_{\gamma}^\tn(\delta_{\alpha\gamma} -
%   \pmv_\alpha\conc_\gamma)(\delta_{\beta\gamma} + \p\ln
%   \textsf{\textsl{f}}_\gamma/\p\ln\conc_\beta)
% \end{equation}
\begin{multline}
  %% Single column style
  \label{eq:13}
  \bn\chempot_\alpha 
  =
  - \chargedens \pmv_\alpha \bn\elpot
  + R \tempr  \sum_{\beta=1}^\tn
  \textsf{\textsl{TDF}}_{\alpha\beta}
  \frac{
    \bn \conc_\beta
  }{
    \conc_\beta
  }
  \\
    + \pmv_\alpha
  \left(
    \left( \lambda
      + \eta/3
    \right)
    \bn
    \left( \bn\vel
    \right)
    + \eta \bn^2 \vel
  \right),
\end{multline}
\noindent where
\begin{equation}
  \label{eq:tdf}
  \textsf{\textsl{TDF}}_{\alpha\beta} =
  \sum_{\gamma=1}^\tn(\delta_{\alpha\gamma} -
  \pmv_\alpha\conc_\gamma)(\delta_{\beta\gamma} + \p\ln
  \textsf{\textsl{f}}_\gamma/\p\ln\conc_\beta)
\end{equation}
is the thermodynamic factor. It is determined by the constitutive
equation
$\bn \ln \conc_\alpha/\conc +\sum_{\beta=1}^\tn(\delta_{\alpha\beta} -
\pmv_\alpha\conc_\beta) \bn
\partial(\massdens\freedens^{\text{int}})/\partial\conc_\beta $, and
differs from the standard form,\cite{Landesfeind2016} as it accounts
for mean steric effects due to nonvanishing molar volumes. Note that
the bulk modulus $\bulkmodalt$, which is not a viable material
parameter in the incompressible limit, disappears in \cref{eq:13}.

\Cref{eq:13} constitutes the complete, mechanically coupled form for
the chemical potentials in our electrolyte model. These thermodynamic
forces obey a viscous Gibbs-Duhem relation,
$\sum_{\alpha=1}^\tn \conc_\alpha (\bn
\chempot_\alpha+F \valence_a
\bn\elpot)$=$ ( \lambda + \eta ) \bn ( \bn\vel) + \eta \bn^2
\vel)$. In equilibrium we can write this with the thermodynamic
pressure defined in \cref{eq:mean_normal_stress} as
$\bn p^{\ce{td}} = -\chargedens\bn\elpot - \bn(\elf\delf)/2$. As
consequence, strong pressure gradients emerge in electrochemical
double-layers.\cite{C7CP08243F}

We use Poisson's equation for the coupling of electric potential and
charge density in the electrolyte, and use the dielectric constant
$\dielrel=1+\suszept$. This closes the set of complete isothermal
equations,
\begin{gather}
  \label{eq:49}
  \chargedens =
  - \dielrel\diel \Delta \elpot,
  \\
    \label{eq:49b}
  \frac{
    \p \chargedens
  }{\p t
  }
  =
  - \bn\Jflux
  - \bn(\chargedens\vel)
  ,
  \\
  \label{eq:49a}
  \frac{
    \p\conc_\alpha
  }{
    \p t
  }
  =
  - \bn \Nflux_\alpha
  -\bn \left(
    \conc_\alpha\vel
  \right),
  \quad \alpha\geq 3,
  \\
  \label{eq:49c}
  \bn \vel
=
- \frac{
	\tilde{\pmv}_2
}{
	F\effval_2
} \bn\Jflux
- \sum_{\alpha=3}^\tn
\tilde{\tilde{\pmv}}_\alpha \bn\Nflux_\alpha.
\end{gather}
\Cref{eq:49a} comprise \tn-2 independent equations for the independent
concentrations $\conc_3,\ldots,\conc_\tn$ (where $\conc_1$ and
$\conc_2$ follow from \cref{eq:8}). Here, we restate the ionic fluxes
and the conduction current density,
\begin{gather}
    \Nflux_\alpha
    =  \frac{    \trans_\alpha  }{    F\effval_\alpha
    }  \Jflux
    - \diffusioncomp_{\alpha\tempr} \bn\tempr
    - \sum_{\beta=3}^\tn
    \diffusioncomp_{\alpha\beta}\bn\effeffpot_\beta, 
  \quad \alpha\geq 3\\
  \Jflux   =
  - \elcond\bn   \varphi  - \upbeta\elcond \bn\tempr
  -  \frac{  \elcond  }{  F  }  \sum_{\beta=3}^\tn
  \frac{  \trans_\beta  }{  \effval_\beta  }  \bn \effeffpot_\beta, 
\end{gather}
which are subject to the form of the chemical potentials following
from the modeled free energy density \cref{eq:13}. Note that
$\varphi = \elpot + \effpot_2 / F \effval_2$ is the reduced
electrochemical potential introduced in
\cref{eq:electrochempoteff}. Volume-expansion $\bn\vel$ is determined
by \cref{eq:2}. Alternatively, the complete set of transport equations
can be cast into matrix-form (see \cref*{SIsec:supp_matr-form-transp}).

In \cref*{SIsec:thermal-aspects}, we complete the set of transport
equations by deriving the equation for temperature (``heat
equation''),
\begin{multline}
  %% Two column style
\label{eq:64}
\textsf{\textsl{{C}}}\dot{\tempr}
=
\mathcal{R}
- \upmu_{\ce{T}} \Jflux\bn\tempr
- \tempr \tilde{\upbeta}\bn\Jflux
+\frac{\tempr}{F }\bn \tilde{\textsf{\textsl{D}}}_\chargedens
\bn \chargedens
+ \tempr \sum_{\gamma=3}^\tn \tilde{\textsf{\textsl{D}}}_\gamma
\bn \conc_\gamma
\\
+\tempr\bn\tilde{\uplambda}\bn \tempr.
\end{multline}
% \begin{equation}
%   %% Single column style
% \label{eq:64}
% \textsf{\textsl{{C}}}\dot{\tempr}
% =
% \mathcal{R}
% - \upmu_{\ce{T}} \Jflux\bn\tempr
% - \tempr \tilde{\upbeta}\bn\Jflux
% +\frac{\tempr}{F }\bn \tilde{\textsf{\textsl{D}}}_\chargedens
% \bn \chargedens
% + \tempr \sum_{\gamma=3}^\tn \tilde{\textsf{\textsl{D}}}_\gamma
% \bn \conc_\gamma
% +\tempr\bn\tilde{\uplambda}\bn \tempr.
% \end{equation}

\section*{Simulation of Zinc Ion Battery Cell}
\label{sec:model-zincpr-blue}

Secondary zinc-ion batteries (ZIB) are an emerging technology for safe
and low-cost energy
storage.\cite{Clark2020b,xu2012energetic,novelznair} Here, we model a
secondary ZIB with IL-water mixture as electrolyte, based on the
experimental work described in \citenum{Liu2017}. Comparison with
these experimental results serves as validation for our transport
theory. First, we sketch the cell set-up and describe our modeling
equations. Second, we present our simulation results.

\begin{figure}[htb!]
  \centering
  \includegraphics[width=0.4\textwidth]{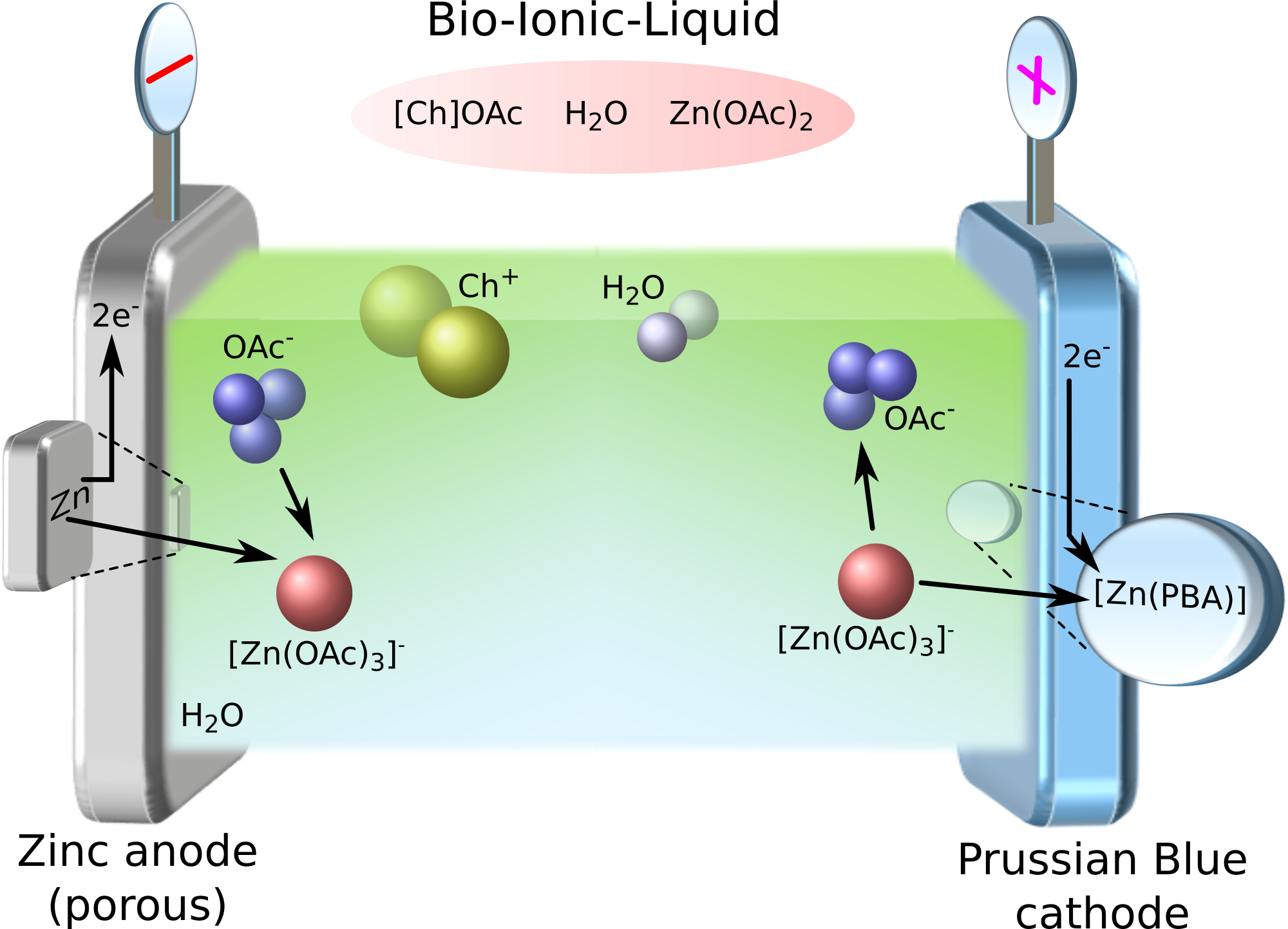}
  \caption{Scheme of the simulated zinc-ion battery.}
  \label{fig:scheme_cell}
\end{figure}

The electrodes consist of a porous zinc-anode (zinc powder) and a
Prussian-blue-analogue (PBA) cathode (\ce{FeFe(CN)_6}). The employed
electrolyte is composed of a mixture of choline acetate (\choac) with
30 wt \% water into which 1M of zinc acetate (\znoac) is
dissolved. Despite the significant amount of water, the \choac+water
mixture can (still) be denoted ``IL with water'', and may thus be
viewed as highly concentrated.\cite{Liu2015,endres_prussian} This
terminology is motivated by the large mass fraction of the salt,
$(\massdens_3^0+\massdens_4^0) / \massdens$, see
\cref{tab:elyte_comp_init}. The ZIB is illustrated in
\cref{fig:scheme_cell}.

\subsection*{Model Equations}
\label{sec:model}
We neglect hydration of the salt-ions by water and assume a complete
dissociation of the salt $\choac \ce{<=>} \chion + \oacion$. In
principle, various different ionic zinc-complexes may form under
bulk-reactions, de\-pen\-ding on the pH-value of the
solution.\cite{clark2017rational,Clark2019} However, such a detailed
investigation lies beyond the scope of this paper. Thus, we assume
complete complexation of the zinc-salt according to
$\znoac + \oacion \ce{<=>} \znoacion$.\cite{DEVREESE2013788} We model
the electrolyte as quaternary mixture of the three charged
constituents $\chion, \oacion, \znoacion$, and
water.\cite{Williams1996,riddick1970organic,fleming1960953} As
reference, we designate water, thus
$\effval_\alpha = \valence_\alpha$. This choice corresponds to
transport theories developed for aqueous solutions, based on water as
solvent.\cite{arnulfbs2}
\begin{table}[!b]
  \begin{ruledtabular}
    \begin{tabular}{lrlll}
        Species
      &
        $\effval_\alpha$
      &
        $\conc_\alpha^0$ / $\si{\mol\per\meter\cubed}$
      &
        $\massdens_\alpha^0$ / $\massdens$
      &
        $\conc^0_\alpha\pmv_\alpha$
      \\
      \midrule
        $\alpha$=1: \ce{H_2O} (designated)
      &
        \tablenum[table-space-text-pre=-,table-format= -1.0]{0}
      &
        \tablenum[table-format= 2.2e2]{19.43e3}
      &
        \tablenum[table-format= 1.2]{0.26}
      &
        \tablenum[table-format= 1.2]{0.35}
      \\
        $\alpha$=2: \chion
      &
        \tablenum[table-space-text-pre=-,table-format= -1.0]{1}
      &
        \tablenum[table-format= 2.2e2]{5.00e3}
      &
        \tablenum[table-format= 1.2]{0.39}
      &
        \tablenum[table-format= 1.2]{0.35}
      \\
        $\alpha$=3: 
      \oacion
      &
        \tablenum[table-space-text-pre=-,table-format= -1.0]{-1}
      &
        \tablenum[table-format= 2.2e2]{4.00e3}
      &
        \tablenum[table-format= 1.2]{0.11}
      &
        \tablenum[table-format= 1.2]{0.22}
      \\
        $\alpha$=4: 
      \znoacion
      &
        \tablenum[table-space-text-pre=-,table-format= -1.0]{-1}
      &
        \tablenum[table-format= 2.2e2]{1.00e3}
      &
        \tablenum[table-format= 1.2]{0.18}
      &
        \tablenum[table-format= 1.2]{0.08}
    \end{tabular}
  \end{ruledtabular}
  \caption{Species assignment and initial electrolyte
    composition. Since water is the designated species,
    $\effval_\alpha=\valence_\alpha$. The columns show initial
    concentrations $\conc_\alpha^0$, mass fractions
    $\massdens_\alpha^0$ / $\massdens$, and volume fractions
    $\conc^0_\alpha\pmv_\alpha$.}
  \label{tab:elyte_comp_init}
\end{table}

\begin{figure}[!t]
  \centering
  \includegraphics[width=8.255cm]{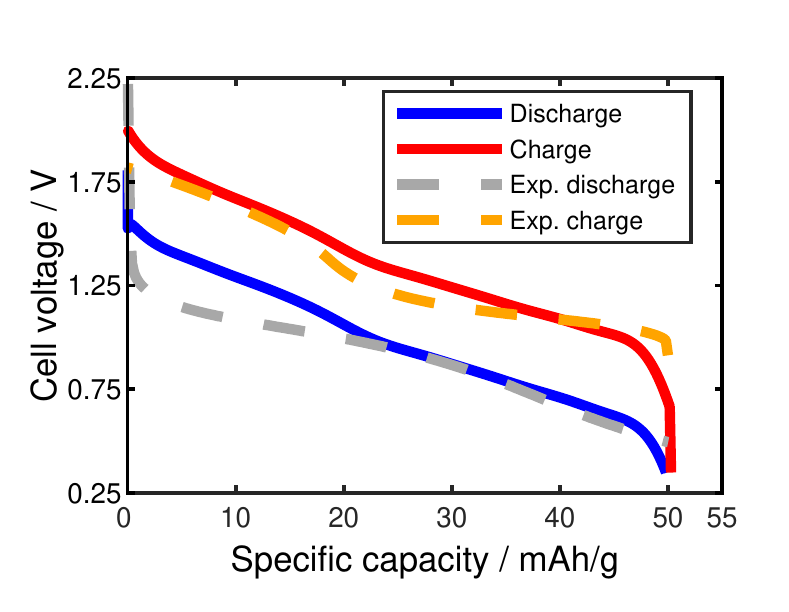}
  \caption{Comparison of simulated and measured\cite{Liu2017} cell
    voltage for a galvanostatic charge-discharge cycle of the ZIB with
    current density
    $I=\SI{0.1}{\milli\ampere\per\centi\meter\squared}$.}
  \label{fig:cycling}
\end{figure}

Since length scales above some nanometers suffice for the description
of batteries,\cite{arnulfbs2} we can safely assume our system to be
electroneutral, \textit{viz.} $\chargedens = 0$.\cite{Newman2004} As a
consequence, the charge density is not a free variable and we solve
$\p_t \chargedens=0$ in \cref{eq:49b} for $\bn\Jflux$ instead of
Poisson's \cref{eq:49}. As thermal aspects play a minor role in
elementary battery cells,\cite{arnulfbs} we assume thermal equilibrium
at room temperature, \textit{i.e.}, $\bn\tempr=0$. Hence, the
temperature appears only as constant a parameter. Here, we set the
activity coefficients to $\textsf{\textsl{f}}_\alpha=1$ and neglect
non-ideal interactions $\freedens^{\ce{int}}=0$. Furthermore, we
assume constant viscosity-coefficients
($\eta=\SI{25.3}{\milli\pascal\second}$ and
$\lambda=0$)\cite{IoLiTecIonicLiquidsTechnologiesGmbh2012} in the
chemical potentials.

The only independent specific concentrations are $\conc_{3}$ and
$\conc_{4}$, which determine $\conc_{1}$ and $\conc_2$ via
\cref{eq:8}, subject to the condition $\chargedens=0$.

Due to the porous cell structure, the set of transport equations must
be modified by means of volume-averaging, as described by porous
electrode theory.\cite{Schmitt2020} Volume-averaging over liquid and
solid phases implies volumetric ($\upvarepsilon$) and dynamic
($\upvarepsilon^\upbeta$) transport-corrections, where
$\upvarepsilon=V^{l}/V$ is the liquid-phase volume-fraction or
porosity. $\upbeta$ is the Bruggemann-coefficient, which depends on
the microstructure,
\begin{gather}
  \label{eq:29}
  \frac{
    \p
  }{
    \p t}
  \left(
    \upvarepsilon \conc_\alpha
  \right)|_{\alpha=3,4}
  = -\bn \left(
    \upvarepsilon
    \conc_\alpha\vel
  \right)
  - \bn
  \left(
    \upvarepsilon^\upbeta
    \Nflux_\alpha
  \right)
  +  r_\alpha,
  \\
  0=
  \bn\left(
    \upvarepsilon^\upbeta \Jflux
  \right)
  = -\bn \left(
    \upvarepsilon
    \chargedens\vel
  \right)
  + \sum_{\alpha=1}^4
  F\valence_\alpha r_\alpha,
  \\
  \bn
  \left(
    \upvarepsilon
    \vel
  \right)
  = \sum_{\alpha=1}^4 \pmv_\alpha
  r_\alpha
  - \frac{
    \tilde{\pmv}_2
  }{
    F\effval_2
  } \bn
  \left( \upvarepsilon^\upbeta\Jflux\right)
  - \sum_{\alpha=3}^4
  \tilde{\tilde{\pmv}}_\alpha \bn
  \left(
    \upvarepsilon^\upbeta \Nflux_\alpha
  \right). \label{eq:740}
\end{gather}
Reactions occur only at the electrode surfaces and appear as
source-terms in our model equations, defined as
\begin{equation}
  \label{eq:54}
   r_\alpha
  =
  \sum_{k, \varGamma}
  a^{\varGamma}
  \nu_{k;\alpha}^{\varGamma}
  i^{\varGamma}.
\end{equation}
This sum includes all reactions $k$ at all electrode-interfaces
$\varGamma$, involving species $\alpha$. Here,
$\nu_{k;\alpha}^{\varGamma}$ are the stoichiometries of the reactions,
and the specific surfaces $a^{\varGamma}$ measure the
surface-to-volume ratio of the electrode $\varGamma$. The specific
surface-reaction-rate $i^{\varGamma}$ constitutes interface-conditions
between the solid and liquid phase, and depends upon the reaction
overpotential. We model this quantity using a
Butler-Vollmer-Ansatz,\cite{Latz2013358} see
\cref*{SIsec:suppformchemreact}.

The half cell ractions occuring at the \ce{Zn}-anode and the
PBA-cathode (\ce{ FeFe(CN)_6}) are (see \cref*{SIsec:suppelectrcomp})
\begin{gather}
%%Two column style
  \label{eq:halfcellreactionzincanode}
  \ce{Zn + 3OAc^-} \ce{<=> 2e^- + [Zn(OAc)_3]^-},
  \\
  \label{eq:halfcellreactionpbacathode}
  \begin{split}
    \ce{[Zn(OAc)_3]^-} + \ce{2e^-}+ \ce{ FeFe(CN)_6}
    &\ce{<=>}
    \\
    \ce{3OAc^-} &+ \ce{[ZnFeFe(CN)_6]}.
  \end{split}
\end{gather}
% \begin{gather}
%   %% Single column style
%   \label{eq:halfcellreactionzincanode}
%   \ce{Zn + 3OAc^-} \ce{<=> 2e^- + [Zn(OAc)_3]^-},
%   \\
%   \label{eq:halfcellreactionpbacathode}
%     \ce{[Zn(OAc)_3]^-} + \ce{2e^-}+ \ce{ FeFe(CN)_6}
%     \ce{<=>}
%     \ce{3OAc^-} + \ce{[ZnFeFe(CN)_6]}.
% \end{gather}
Thus, zinc dissolves from the anode and intercalates into the
cathode (and vice versa). We neglect solid-state diffusion and model
the cathodic reactions as deposition-processes (see 
\cref*{SIsec:suppsolidphaseequat}).

\subsection*{Simulation Results}
\label{sec:simulation}
\begin{figure}[!t]
  \centering
  \includegraphics[width=8.255cm]{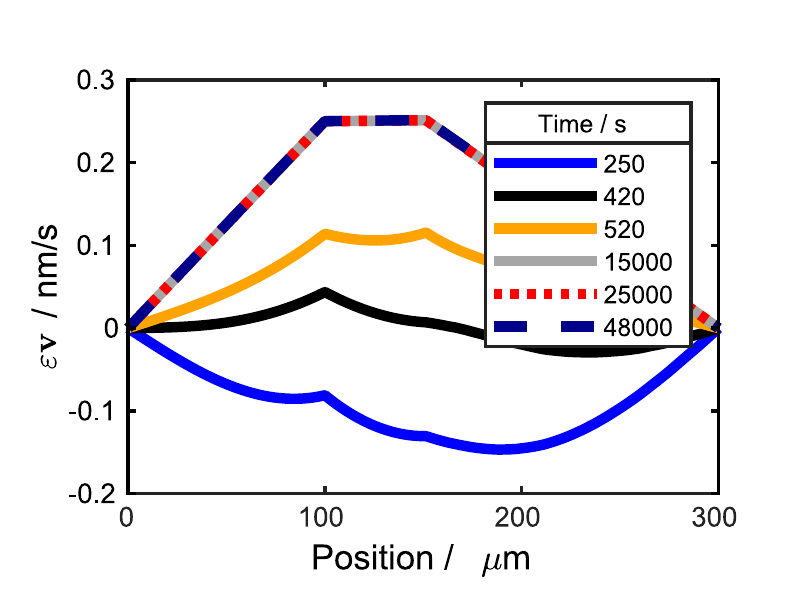}
  \caption{Volume-averaged convection velocities at the designated
    moments (see \cref*{SIfig:cv_overview}) during discharge of the
    ZIB with current density
    $I=\SI{0.1}{\milli\ampere\per\centi\meter\squared}$.}
  \label{fig:advection}
\end{figure}
In this section, we present the simulation results of the as-modeled
ZIB and compare them with the experimental results (we specify our
numerical methods in \cref*{SIsec:suppcompdetamatl}).\cite{Liu2017}
In this way, we study the spatial profile and temporal evolution of
concentrations and convection velocity. We highlight the relevance of
convection for cell performance and compare possible definitions of
transference numbers.

\subsubsection*{Competition of Diffusion, Migration, and Convection}
First, we validate our model. For this purpose, we simulate the
galvanostatic discharge and charge of the ZIB by applying a moderate
current density
($I=\SI{0.1}{\milli\ampere\per\centi\meter\squared}$). We compare our
simulation results with experimental observations in
\cref{fig:cycling}. Apparently, the specific capacities and cell
voltages obtained from simulation are in good agreement with the
experimental values. However, the discharge profile obtained from
experiment exhibits two discharge phases with a transition at
approximately $\SI{20}{\milli\ampere\hour\per\gram}$. Endres et
al. \cite{Liu2017} attribute this effect to two different
electro-reactivities, stemming from low-/ and high-spin states of
\ce{Fe(III)} in the PBA. Because such atomistic processes lie beyond
the scope of our model in this paper, this transition is not found in
simulations.

Our model provides a complete description of the electrolyte
dynamics. In the following, we discuss the ion dynamics and their
influence on the overall cell performance in order to understand the
interplay between the different transport mechanisms (migration,
diffusion, and convection), and the reactions at the electrode
surfaces.

For a proper discussion of the electrolyte evolution during discharge
of the battery, we designate some characteristic
moments. \Cref*{SIfig:cv_overview} shows the discharge-profile of the
cell-voltage and the designated moments. We set one focus on processes
occuring during the initial discharge-phase
($t=\SI{250}{\second},\SI{420}{\second},\SI{520}{\second}$). Furthermore,
we designate two intermediate moments, and, finally, the moment of
complete discharge of the cell. The significance of the designated
times becomes apparent below when evaluating the
electrolyte-quantities $\elpot$, $\vel$, and $\conc_\alpha$ at these
moments.

\begin{figure}[!t]
  \centering
  \includegraphics[width=8.255cm]{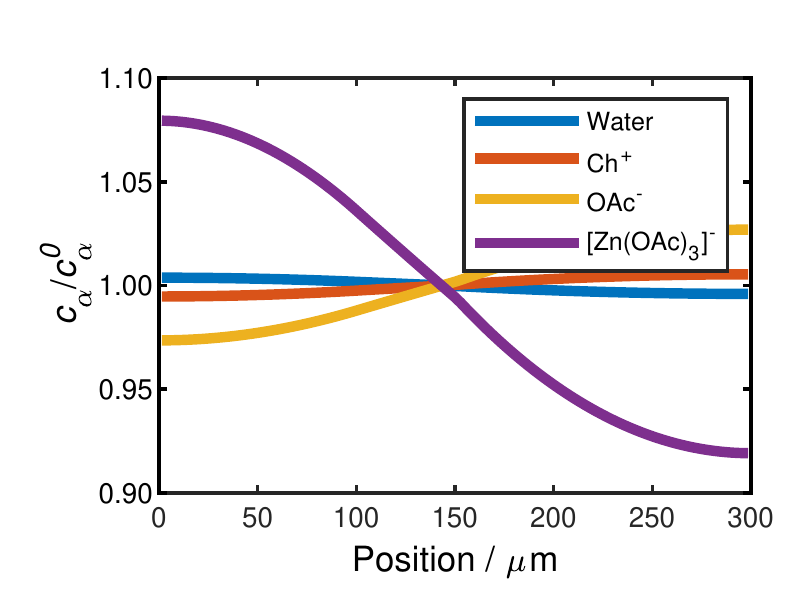}
  \caption{Typical concentration profiles for the quaternary
    electrolyte-composition at the end of discharge with discharge
    current density 
    $I=\SI{0.1}{\milli\ampere\per\centi\meter\squared}$. The initial
    electrolyte composition, $\conc_\alpha^0$, is tabulated in
    \cref{tab:elyte_comp_init}}
  \label{fig:c_typical}
\end{figure}

\Cref{fig:advection} shows the temporal evolution of the convection
velocity profile. Apparently, the direction of the motion of the
center-of-mass switches from towards the anode (negative convection),
to towards the cathode (positive convection) right after the initial
phase, and, then, the profile quickly becomes quasi-stationary. As
discussed below, $\vel$ is not dominant in our system. In the
supplementary, we observe a similar behavior for the ionic
concentrations (see \cref*{SIfig:cv_overview}c)-f)). Initially (first
three designated moments), concentration-differentials grow throughout
the cell, followed by quasi-stationary profiles during the rest of the
discharge. This suggests that the dynamics of the concentration and
the convection profile arise from a common initiation-mechanism. Due
to the interplay of multiple transport mechanisms, i.e., diffusion,
migration, and convection, the dynamics of the concentrations and of
the convection velocity are coupled.

We explain these observations with the push of the electrolyte out of
its initial equilibrium-state by the application of the discharge
current through the reactions at the electrodes. As the system relaxes
towards the new stationary state under galvanostatic discharge, the
concentrations approach stationary profiles. Spontaneous electrode
reactions directly influence the profile of the ion concentrations
near the electrodes. In addition, concentration gradients are built
up, due to convective transport of zinc from anode to cathode. As zinc
is dissolved as complex $\znoacion$, a net loss of available
electrolyte volume occurs at the cathode, and a net gain of
electrolyte volume at the anode. This pushes electrolyte from cathode
to anode according to \cref{eq:740} and results in negative convection
velocities. We give a detailed analysis of the electrolyte-dynamics in
the \supi{} (see \cref*{SIsec:elyte_dynamics}).

The evolution of the electric potential in the electrolyte is shown in
\Cref*{SIfig:cv_overview}b). As expected, the cathode is more
electronegative than the anode
($\Delta\elpot^{\ce{cell}}\approx\SI{-0.1}{\milli\volt}$) at all times
during discharge. This implies a migrational pull of the anionic
species towards the anode, and of the \chion-ions towards the
cathode. Thus, migration hinders cell operation, which depends on
negative \znoacion-ions. As a consequence, diffusion and convection
must overcompensate the electric forces in the electrolyte to sustain
cell operation.

\begin{figure}[!t]
  \centering
  \includegraphics[width=8.255cm]{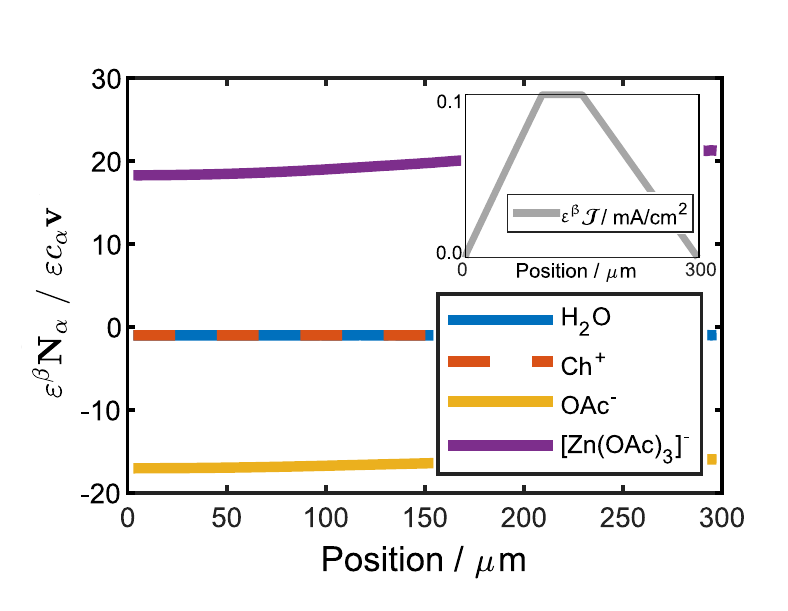}
  \caption{Convection contribution to ionic mass fluxes at the
    end of discharge. The inset shows the conduction current
    density $\upvarepsilon^\upbeta \Jflux$ at end of discharge
    with current density 
    $I=\SI{0.1}{\milli\ampere\per\centi\meter\squared}$.}
  \label{fig:N_vs_advec}
\end{figure}

\Cref{fig:c_typical} shows normalized, quasi-stationary
con\-cen\-tra\-tion-pro\-files of all electrolyte species at
end-of-discharge. Interestingly, the gradients of the two anionic
species (\oacion{} and \znoacion) have opposite orientations. The
profile for the positively charged \chion-ions moderately increases at
the more electronegative cathode. Likewise, water shows a small
gradient, despite being neutral ($\effval_{\ce{H_2O}}=0$ in this
reference-frame) and thus not being susceptible to migration.

Most importantly, the large concentration gradient of the
\znoacion-ions is necessary to overcome the migration pull. This is
mandatory to maintain cell operation, as zinc is intercalated into the
cathodic PBA-structure.  \oacion-ions have to move back to the anode,
which is realized by a small concentration gradient and the migration
pull. This motion of bulky \oacion-ions induces the positive
convection velocity during the stationary phase of galvanostatic
discharge seen in \cref{fig:advection} (see \cref{eq:740}).

\Cref{fig:N_vs_advec} illustrates the relevance of convective
transport for the electrolyte-species. The ratio
$\upvarepsilon^\upbeta \Nflux_\alpha/\upvarepsilon \conc_\alpha \vel$
shows the relation between flux densities within the center-of-mass
system and the center-of-mass velocities. This ratio is larger than
one (roughly one), if convection is negligible (dominant) in transport
of the respective species. Apparently, convective flux contributions
are negligible for the two negative species, whereas convection plays
a significant role for the dynamics of water and \chion-ions. Thus,
convection is important for those species that do not contribute to
the half-cell reactions. We infer from the sign of the ratio
$\upvarepsilon^\upbeta \Nflux_\alpha/\upvarepsilon \conc_\alpha \vel$
that \oacion-ions move towards the anode, whereas \znoacion-ions moves
towards the cathode. This confirms our previous finding for the
\znoacion-ions that diffusion overcompensates migration.

\subsubsection*{Limiting Discharge Currents}
Now, we investigate the power limiting mechanisms of this ZIB with IL
electrolyte. For this purpose, we simulate the cell discharge at high
current densities.

\begin{figure}[!tb]
  \centering
  \includegraphics[width=8.255cm]{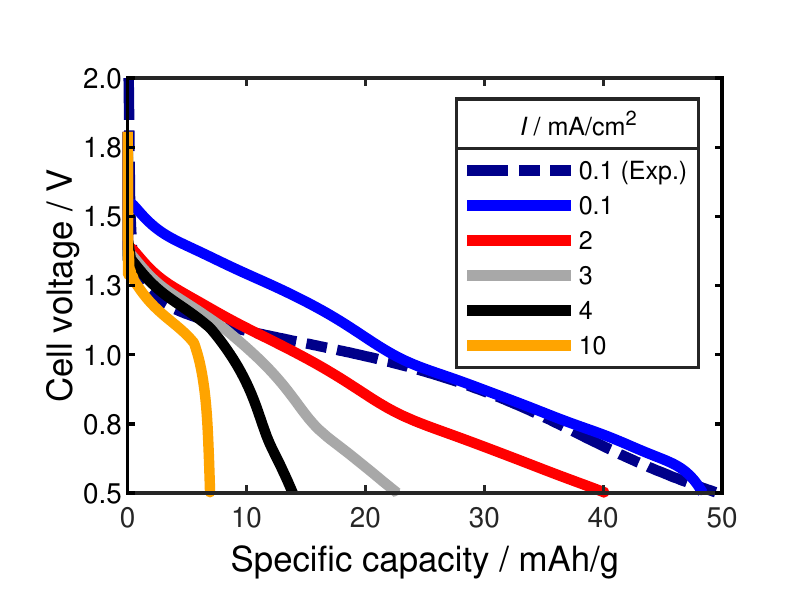}
  \caption{Cell voltage during discharge for various 
    current densities.}
  \label{fig:crnts_loop_expsim}
\end{figure}
\Cref{fig:crnts_loop_expsim} illustrates the impact of increased
discharge currents on the overall cell performance. The shape of the
discharge profile is preserved for moderate discharge current
densities (up to $I=\SI{2}{\milli\ampere\per\centi\meter\squared}$),
despite being shifted to smaller discharged capacities. At higher
current densities, a strong capacity fade is observed and the
discharge profiles exhibit steep voltage drops.

These voltage drops suggest that diffusion becomes too slow to supply
the cathodic interfacial reaction mechanism with sufficient amount of
\znoacion-salt. Such a behavior is typical for ILs, which are often
highly viscous.\cite{C3TA15273A}

\begin{figure}[htb!]
  \centering
  \includegraphics[width=8.255cm]{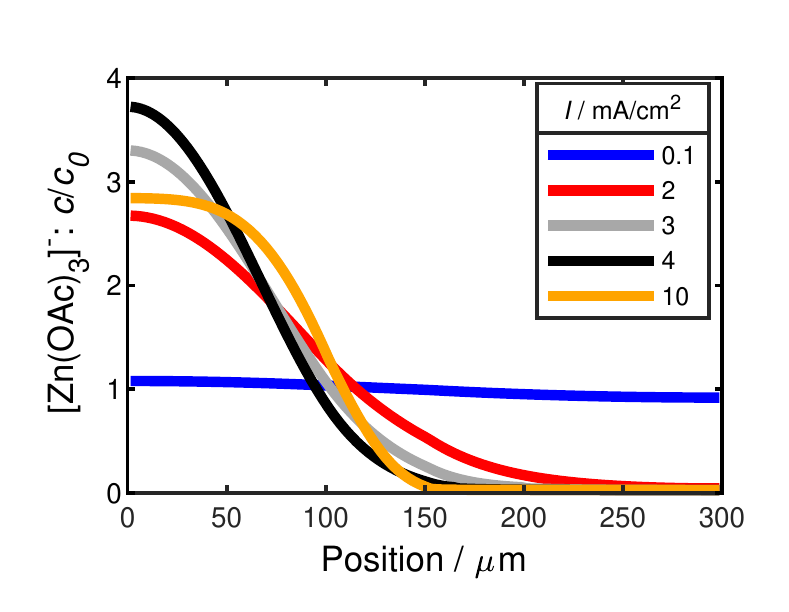}
  \caption{Concentration profiles of \znoacion{} at end of discharge
    state for various current densities.}
  \label{fig:depletion_zncomplex}
\end{figure}

This explanation is confirmed by \cref{fig:depletion_zncomplex}, in
which the concentration profiles of the \znoacion{}-ions for the
different discharge currents at the end of discharge is shown. We
observe steep concentration gradients throughout the cell. The salt
concentration at the cathode decreases with increasing discharge
currents, and, finally, \znoacion{}-ions get depleted.

At intermediate current densities, we observe in
\cref{fig:depletion_zncomplex} that the \znoacion{} concentration at
the anode increases with discharge current density. Interestingly, it
starts to decrease for very high current densities. In this case, the
cell fails so quickly due to diffusion limitation that a
quasi-stationary concentration-profile cannot establish.

\subsubsection*{Discussion of Transference Numbers}
Finally, we demonstrate the consistency of our transport theory. In
our model, we choose two reference species because of momentum and
charge conservation. The model predictions should be independent of
this choice. Here, we compare simulations with different reference
species.

In the theory-chapter, we showed that all $\tn(\tn{+}1)/2$
tran\-sport-parameters, appearing in a \tn-component
elec\-tro\-lyte-mixture, follow from the Onsager matrix
$\boldsymbol{\mathcal{L}}$, see
\cref{eq:def_elcond,eq:def_seebeck,eq:def_heatcond,eq:def_transferencenumber,eq:D_aT}. Although
the designated species does not appear explicitly,
$\boldsymbol{\mathcal{L}}$ comprises the complete set of
inter-species-correlations, including correlations involving the
designated species. This becomes apparent in the clo\-sure-relation
for the independent fluxes $\Nflux_\alpha|_{\alpha\geq 2}$,
\cref{eq:27}, which implicitly determines $\Nflux_1$. Thus, the
Onsager-matrices with respect to different designated species are
mutually coupled, and cannot be stated independently from each
other. In \cref*{SIsec:supp_transf-refer-spec} we show that simple
conversion relations exist, which allow to transfer between the
different choices of reference species.

In \cref*{SIsec:suppcompresultsdiff} we prove consistency of our
framework. \Cref*{SIfig:delta_compare_A} shows a comparison of the
discharge curves and the final concentration profiles of the
\znoacion-ions, obtained by simulations using the three different sets
of reference species comprised in
\cref{tab:alternative_speciesassign}. Apparently, the deviations lie
within numerical accuracy (relative error
$\approx\mathcal{O}(10^{-6})$). This is consistent, since both cell
voltage and concentrations must not depend upon the choice of
reference. Thus, they give, up to numerical accuracy, identical
results.

\begin{figure}[htb!]
  \centering
  \includegraphics[width=8.255cm]{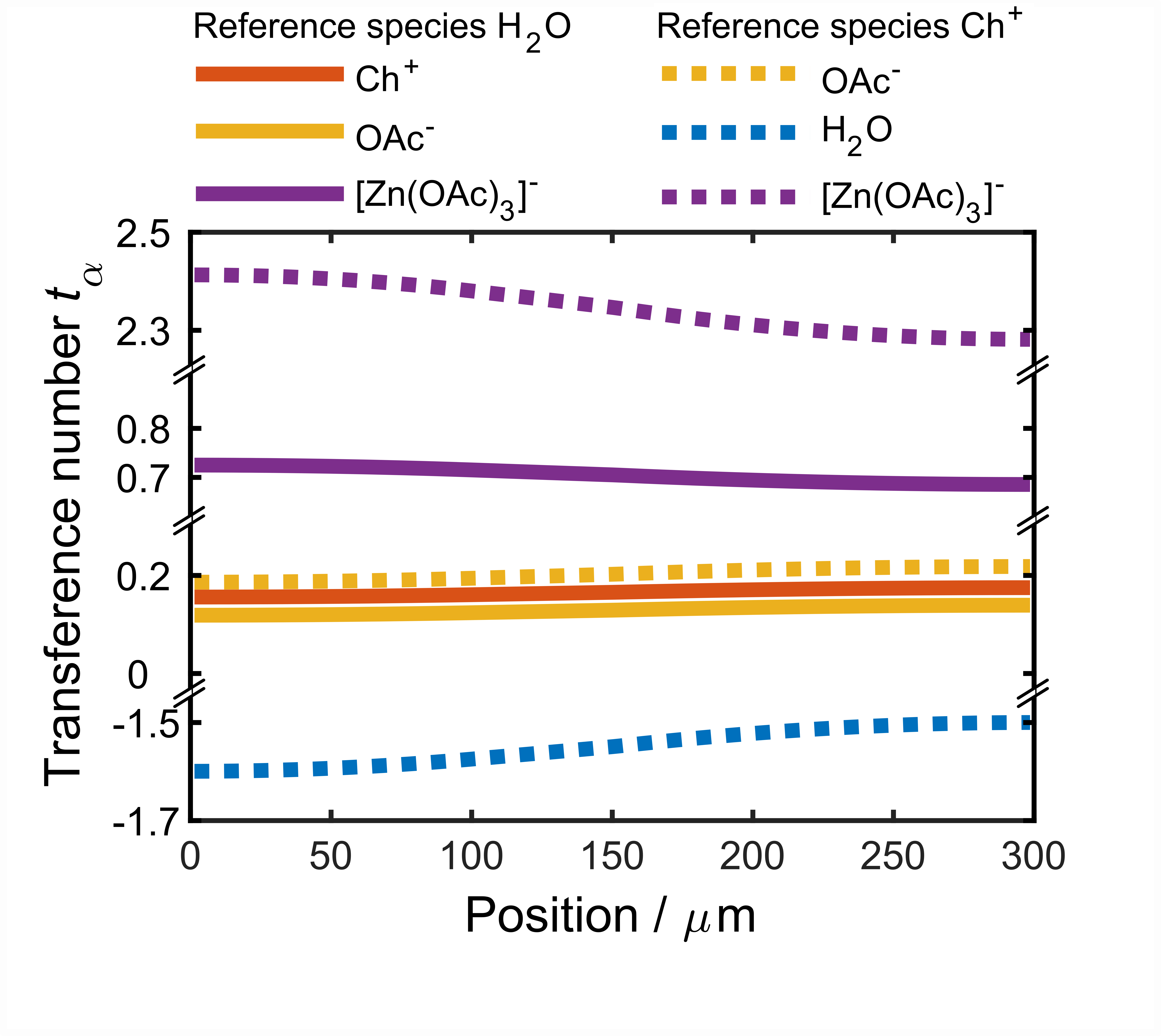}
  \caption{Transference numbers at end of discharge for the reference
    species $\ce{H_2O}$, and $\chion$ (discharge current density
    $I=\SI{0.1}{\milli\ampere\per\centi\meter\squared}$).}
  \label{fig:t_typical}
\end{figure}

\begin{table}[!b]
  \begin{ruledtabular}
    \begin{tabular}{llll}
      Reference
      &
        Species
      &
        $\effval_\alpha$
      &
        $\trans_\alpha$
      \\
      \hline
      \multirow{4}{*}{\STAB{\rotatebox[origin=c]{90}{Water}}}
      &
        Water
      &
        \tablenum[table-space-text-pre=-,table-format= -1.3]{0}
      &
        \textit{n.d.}
      \\
      &
        \chion
      &
        \tablenum[table-space-text-pre=-,table-format= -1.3]{1}
      &
        \tablenum[table-space-text-pre=-,table-format= -1.3]{0.166}
      \\
      &
        \oacion
      &
        \tablenum[table-space-text-pre=-,table-format=
        -1.3]{-1}
      &
        \tablenum[table-space-text-pre=-,table-format= -1.3]{0.129}
      \\
      &
        \znoacion
      &
        \tablenum[table-space-text-pre=-,table-format= -1.3]{-1}
      &
        \tablenum[table-space-text-pre=-,table-format= -1.3]{0.705}
      \\
      \hline
      \multirow{4}{*}{\STAB{\rotatebox[origin=c]{90}{\chion}}}
            &
        Water
      &
        \tablenum[table-space-text-pre=-,table-format= -1.3]{-0.170}
      &
        \tablenum[table-space-text-pre=-,table-format= -1.3]{-1.549}
      \\
      &
        \chion
      &
        \tablenum[table-space-text-pre=-,table-format= -1.3]{0}
      &
        \textit{n.d.}
      \\
      &
        \oacion
      &
        \tablenum[table-space-text-pre=-,table-format= -1.3]{-1.570}
      &
        \tablenum[table-space-text-pre=-,table-format= -1.3]{0.203}
      \\
      &
        \znoacion
      &
        \tablenum[table-space-text-pre=-,table-format= -1.3]{-3.330}
      &
        \tablenum[table-space-text-pre=-,table-format= -1.3]{2.346}
      \\
      \hline
      \multirow{5}{*}{\STAB{\rotatebox[origin=c]{90}{\znoacion}}}
      &
        Water
      &
        \tablenum[table-space-text-pre=-,table-format= -1.3]{0.074}
      &
        \tablenum[table-space-text-pre=-,table-format= -1.3]{0.665}
      \\
      &
        \chion
      &
        % \tablenum[table-space-text-pre=-,table-format= -1.2e3]{1.430}
        \tablenum[table-space-text-pre=-,table-format= -1.3]{1.430}
      &
        \tablenum[table-space-text-pre=-,table-format= -1.3]{0.237}
      \\
      &
        % $\alpha=3$:
        \oacion
      &
        \tablenum[table-space-text-pre=-,table-format= -1.3]{-0.757}
      &
        \tablenum[table-space-text-pre=-,table-format= -1.3]{0.098}
      \\
      &
        \znoacion
      &
        \tablenum[table-space-text-pre=-,table-format= -1.3]{0}
      &
        \textit{n.d.}
      \\
      \\
    \end{tabular}
  \end{ruledtabular}
  \caption{Results for the transference-numbers obtained from using
    different reference-frames. The numbers were obtained from 
    spatially-averaging over the cell. No transference-numbers for the
    designated species ($\alpha=1$) exist relative to the
    center-of-mass motion (\textit{n.d.}, not defined). In particular,
    $\sum_{\alpha=2}^\tn\trans_\alpha=1$ in each frame.}
  \label{tab:alternative_speciesassign}
\end{table}
In contrast to the electric conductivity, some transport parameters,
\emph{e.g.}, transference numbers, depend on the choice of reference
species. \Cref{fig:t_typical} shows the transference numbers at the
end of discharge for two different reference species. In addition,
\cref{tab:alternative_speciesassign} summarizes spatially averaged
values of the transference numbers at end of discharge, for three
different sets of reference species. In quaternary electrolytes, only
two transference numbers are independent and only three transference
numbers are well-defined.

We observe in \cref{fig:t_typical} that sign and magnitude of the
transference numbers depends on the reference species chosen due to
momentum conservation. Interestingly, $\trans_{\oacion}$ is rather
similar for both reference species $\ce{H_2O}$, and $\chion$,
respectively. A significant discrepancy is observed for
$\trans_{\znoacion}$. When $\chion$ as charged species is designated,
water acquires an effective charge, and thus contributes to the
electric current. In particular, $\trans_{\ce{H_2O}}$ is negative in
the $\chion$-frame.

Transference numbers are particularly intuitive for the neu\-tral
re\-fe\-rence species $\ce{H_2O}$. In this setting, the signs of
transference numbers endorse our interpretation of the over\-all
electrolyte dynamics, discussed above. % At end of discharge,
% \cref{tab:alternative_speciesassign} and the inset in
% \cref{fig:N_vs_advec} imply that
% $\trans_{\znoacion}\Jflux/\effval_{\znoacion}{<}0$, whereas
% \cref{fig:advection,fig:N_vs_advec} imply
% $\Nflux_{\znoacion}{>}0$. Thus, diffusion dominates over migration and
% is the main driving force for cell operation (see
% \cref{eq:alternative23}).

\Cref{tab:alternative_speciesassign} and the inset in
\cref{fig:N_vs_advec} imply that at end of discharge
$\trans_{\znoacion}\Jflux/\effval_{\znoacion}{<}0$, whereas
\cref{fig:advection,fig:N_vs_advec} imply
$\Nflux_{\znoacion}{>}0$. Thus, diffusion dominates over migration and
is the main driving force for cell operation (see
\cref{eq:alternative23}).

The relations between the numerical results for the transference
numbers obtained in the three different reference-frames agree with
the analytical predictions, see %\cref*{SIeq:ex_trns_ex_ind}.
\cref*{SIeq:ex_ch_trns_ex_ind_1,SIeq:ex_ch_trns_ex_ind_2,SIeq:ex_ch_trns_ex_dep,SIeq:ex_zn_trns_ex_ind_1,SIeq:ex_zn_trns_ex_ind_2,SIeq:ex_zn_trns_ex_dep}.

Since, there is an ongoing debate regarding the sign and magnitude of
transference numbers in concentrated electrolytes and ionic
liquids,\cite{C7CP08580J,C8CP02595A} we shall briefly comment on this
topic here.

Various, differing definitions for transport parameters exist in the
literature, which bears the potential for confusion when comparing
results from different authors. Thus, a complete characterization
should state all transference numbers and make their definitions
clear. In our theory, we define transference numbers and diffusion
coefficients relative to the bulk-motion of the center-of-mass. As it
was shown earlier for molten salts,\cite{ratkje1993transference} such
an internal description predicts that transference numbers can take up
any real value if the Onsager matrix contains off-diagonal terms. Some
works in the literature perform a less general approach and choose one
specific ionic species as internal
reference,\cite{sinistri1962transport} or choose an external
description using fixed coordinates as
reference.\cite{haase1973deutung} Since each approach bears its own
experimental significance,\cite{sundheim1956transference} we resolve
this ambiguity in \cref*{SIsec:lf-transport}. There, we derive simple
relations, which transform between the different reference-systems.

Of course, the theoretical concepts must be probed by
experiments. However, the experimental determination of transport
parameters in concentrated electrolytes is challenging. In principle,
suitable methods must take into account all inter-species-correlations
and the formation of ionic clusters and
ion-aggregates.\cite{vargas2020dynamic} In particular, applicability
of NMR/PFG-NMR experiments for determination of transport parameters
is limited for concentrated electrolytes, as they (i) neglect
formation of ionic complexes, (ii) derive transference numbers from
diffusion coefficients via Nernst-Einstein relations strictly valid
only for ideal electrolytes and (iii) provide only averaged
values.\cite{C8CP02595A} Recently, the measurement of ionic
electrophoretic mobilities was proposed to overcome these obstacles,
either using non-blocking electrodes,\cite{wohde2016li+} or
experiments based on electrophoretic NMR
(eNMR).\cite{BIELEJEWSKI201417,CS9942300165} Unfortunately, the first
technique does not apply to neat IL, as it is not suitable for
non-metal ions.\cite{C5CP05753A} In contrast, eNMR applies to both,
neat ILs and Li-IL-mixtures.\cite{C7CP08580J} In principle, this
method is evaluated with the external fixed laboratory-frame as
reference. However, this is correct only if convection is negligible
(see \cref*{SIeq:all_frames_trans_cmpr_3}). 

\section*{Discussion}
\label{sec:benefits}

In this section, we discuss the relevance of the derived transport
model. We give a detailed comparison with continuum models previously
presented in the literature, before we quantitatively asses the role
of center-of-mass convection and mechanical forces for a consistent
model of highly concentrated electrolytes.

The so-called Newman-model is standard for the mathematical modeling
of batteries on the
cell-level.\cite{Fuller_1994,Thomas2002,Newman2004} This approach
relies on Stefan-Maxwell theory to relate flux densities and chemical
potentials. The canonical Newman model describes a ternary system
composed of a cation, an anion, and a neutral solvent with three
parameters, diffusivity, conductivity, transference number. It is
referred to as concentrated solution theory, as it takes correlations
between anions and cations into account. However, the dynamics of the
solvent is usually neglected.\cite{Fuller_1994} Monroe et al. extend
the Newman model to locally non-neutral multi-component concentrated
electrolytes.\cite{MONROE2013649} They model flux densities relative
to a designated species-velocity, which serves as convection
velocity. In accordance with this description, the transport
parameters relate to the dynamics of the designated species
(``Hittorf-transference'' numbers). However, this approach does not
take into account momentum conservation and mechanical couplings. We
show below that these become relevant in highly concentrated
electrolytes.
	
These Newman-type models can be generalized to account for the
electrolyte equation of state, \cref{eq:38}
.\cite{Horstmann2012a,LIU2014447} We use it to derive the convection
equation in our framework, see \cref{eq:convectionvel}. Monroe et
al. conclude that the relative magnitude of the molar volumes is a
crucial factor (see also
\cref{eq:est_conv_ternary_sys_anly,eq:flux_estimation_interface,eq:6}).\cite{LIU2014447}
Nevertheless, their theory is evaluated only for ternary systems with
neutral solvent.

A systematic approach for the description of liquid el\-ec\-tro\-ly\-tes
has recently been introduced by Latz and
coworkers.\cite{arnulfbs2,arnulfbs} Their approach uses a
ther\-mo\-dy\-na\-mi\-cally con\-sis\-tent
framework,\cite{grootmazur,Coleman1963} which is based on
thermodynamics principles and balancing laws to derive the general
structure of the transport equations with an Onsager
Ansatz.\cite{landau1959fluid,landau2013electrodynamics} They discuss a
ternary system in comparison to the standard Newman model. Mutual
couplings between the ion-species are evaluated, but the dynamics of
the neutral solvent is neglected in the description of
transport. Furthermore, the electro-mechanical forces following from
momentum conservation are not further evaluated.

In a series of publications,\cite{Dreyer2015,Dreyer2014,Dreyer2018}
Guhlke et al. presented a similar framework. Their theory is developed
using the same underlying rigorous assumptions and accounts for
multi-component mixtures. Furthermore, similar to our approach, all
mechanical couplings are incorporated into the transport
equations. One highlight of this framework is the thermodynamic
description of singular surfaces, which is used for the consistent
description of electrochemical double layers. The role of convection
for such thermodynamically consistent frameworks is discussed by the
authors in great generality in Ref.\citenum{dreyer}. However, similar
to the description of Latz and coworkers, the role of convection is
assumed negligible in the discussion of a ternary electrolyte with
neutral solvent.

We extend the approach of Latz and coworkers, and derive a more
general, and more consistent framework, as we do not restrict the
number of species, their valences, nor neglect transport of neutral
solvent. At the same time, we make sure that the theory intuitively
connects to the standard Newman model by using comparable parameter
definitions. We highlight two advances of our model: First,
center-of-mass convection plays a key role in our theory. Thus, molar
masses appear in the definition of transport parameters, see for
example \cref{eq:63}. Second, we take into account the coupling of
mechanics and transport. Thus, molar volumes appear in the chemical
potentials and affect the transport dynamics of the electrolyte, see
for example \cref{eq:tdf}. Due to the consistent reduction of
independent transport parameters, the abstract form of the general
transport equations simplify when specified to particular systems. For
example, the electric conductivity is the only transport parameter
needed to describe a binary IL. In the following, we quantitatively
discuss the importance of these two fundamental extensions for highly
concentrated electrolytes.

\Cref{eq:2} determines the spatial inhomogeneity of the center-of-mass
velocity. We calculate the ratio of the variation of convective flux
density and the variation of non-convective flux density. In a ternary
electrolyte with neutral solvent ($\valence_1=0$) and binary salt
($\conc_2=\conc_3$, $\valence_2=\valence_3$), it holds
\begin{equation}
  \label{eq:est_conv_ternary_sys_anly}
  \left\lvert
    \frac{\conc_3\cdot \bn\vel}{\bn\Nflux_3}
  \right\rvert
  =
  \left\lvert
    \conc_3\cdot \tilde{\tilde{\pmv}}_3
  \right\rvert
  =
  \left\lvert
    1 - \conc_1\pmv_1 \cdot
    \frac{\massdens}{\massdens_1}
  \right\rvert .
\end{equation}
Thus, the relative mass density of the solvent $\massdens_1/\massdens$
and the relative volume density of the solvent $\conc_1\pmv_1$
determine the relative variation of convection velocity. If the
solvent dominates electrolyte mass and volume, i.e.,
$\massdens_1 \sim \massdens$ and $\conc_1\pmv_1\sim 1$, the convection
velocity is constant. If solvent mass is negligible, i.e.,
$\massdens_1 \ll \massdens$, the variation of the center-of-mass
convective flux outweighs the relative variation of the salt
flux. Thus, we define the term highly concentrated salts based on mass
fractions, e.g., $\massdens_1/\massdens$.

The absolute magnitude of the convection velocity is determined by the
electrochemical reactions at the electrode-electrolyte interfaces. The
non-convective species-fluxes
$\Nflux_\alpha + \conc_\alpha\vel = \conc_\alpha\vel_\alpha$ are
subject to flux boundary conditions,
$\conc_\alpha\vel_\alpha|_\varGamma =
\mathscr{R}^\varGamma\nu^\varGamma_\alpha$, at each
electrode-interface $\varGamma$. Here, we model the reaction
source-terms via interfacial currents as defined by \cref{eq:54} such
that
$\mathscr{R}^{\varGamma} = \int \ce{d}\mathbf{x} \,
r_\alpha/\nu^\varGamma_\alpha$, where $\nu^\varGamma_\alpha$ denotes
the stoichiometry of the reactions. Thus, the relevance of convective
fluxes at the interface is determined by the masses / mass-densities,
and the stoichiometries,
\begin{equation}
  \label{eq:flux_estimation_interface}
  \left\lvert
    \frac{\Nflux_\alpha}{\conc_\alpha\vel}
  \right\rvert
  =
  \left\lvert
    1 - \frac{\vel_\alpha}{\vel}
  \right\rvert
  = 
  \left\lvert
    1 - \frac{\massdens}{\massdens_\alpha}
    \cdot \frac{\mass_\alpha\nu^\varGamma_\alpha}{\sum_{\beta=1}^\tn
      \mass_\beta\nu^\varGamma_\beta}
  \right\rvert
\end{equation}
We conclude that the mass fractions $\massdens_\alpha/\massdens$
determine the relevance of the convective flux density. We apply this
analytic result to the zinc ion electrolyte specified in
\cref{tab:elyte_comp_init}. At the anode, we find for the reacting
species
\begin{gather}
  \label{eq:6}
  \left\lvert
    \frac{\Nflux_\alpha}{\conc_\alpha\vel}
  \right\rvert_{\alpha=\znoacion}
  \approx \num{0.05},
  \quad
  \left\lvert
    \frac{\Nflux_\alpha}{\conc_\alpha\vel}
  \right\rvert_{\alpha=\oacion}
  \approx \num{0.06}.
\end{gather}
These values show that in this electrolyte center-of-mass convection
is relevant, but not dominant. This is in agreement with the mass
fraction of water, see \cref{tab:elyte_comp_init}, which is comparable
to the mass fraction of the salt, but not negligible. Thus, in the
bulk the electrolyte behaves as concentrated electrolyte. Our
simulation results in \cref{fig:N_vs_advec} validate our estimates for
inhomogeneity and magnitude of the center-of mass velocity in
\cref{eq:est_conv_ternary_sys_anly} and
\cref{eq:flux_estimation_interface}, respectively.

We note that the apparent complexity of the electrolyte depends on the
choice of reference species as shown in
\cref{fig:t_typical}. Furthermore, electrolyte composition is
significantly altered at electrified interfaces, where ions can
accumulate.\cite{C7CP08243F} It is the unique strength of our
consistent transport theory that it describes both, the interfacial
and the bulk behavior of concentrated electrolytes independent from
the choice of reference species.

We predict an additional contribution to the thermodynamic factor in
\cref{eq:13,eq:tdf}, based on the consistent coupling of mechanical
forces. For ideal electrolytes ($\textsl{\textsf{f}}_\alpha=1$ for all
species $\alpha$), it holds
\begin{align}
  \label{eq:tdf_ideal}
  \bn\chempot_\alpha^{\textsl{\textsf{mixing}}}\big \rvert_{\textsl{\textsf{f}}_\gamma=0}
  &= RT \sum_{\beta=1}^ \tn \textsl{\textsf{TDF}}_{\alpha\beta}\big
  \rvert_{\textsl{\textsf{f}}_\gamma=0}
  \cdot
  \frac{\bn\conc_\beta}{\conc_\beta}\nonumber\\
  &=
  RT
  \left(
    \frac{\bn\conc_\alpha}{\conc_\alpha}
    + \sum_{\beta\neq\alpha}^ \tn
    \left[ \pmv_\beta - \pmv_\alpha\right]
    \bn\conc_\beta
  \right).
\end{align}
Thus, we find cross-couplings between species with different molar
volumes. We discuss the relevance of species-asymmetry with two
limiting cases, where we assume that the concentrations lie in the
same order of magnitude, i.e., $\conc_\alpha\sim\conc_\beta$. First,
if all species have the same molar volume, \cref{eq:tdf_ideal} reduces
to the standard ideal form, where all species are decoupled,
$\bn\chempot_\alpha^{\textsl{\textsf{mixing}}}\big\rvert_{\textsl{\textsf{f}}_\gamma=0}
= RT\bn\conc_\alpha/\conc_\alpha$. Second, if the first species is
much larger than the others,
$\pmv_1\gg\pmv_\alpha\rvert_{\alpha\geq 2}$, then the forces on the
small species effectively decouple,
$\bn\chempot_\alpha^{\textsl{\textsf{mixing}}}\rvert_{\alpha\geq 2} =
RT\cdot \bn\conc_\alpha/\conc_\alpha$, whereas the designated species
experiences only inter-species correlations
$\bn\chempot_1^{\textsl{\textsf{mixing}}} = -RT\pmv_1 \sum_{\beta\neq
  1}^\tn \bn\conc_\beta$. This analyis exemplifies the relevance of
the consistent coupling of transport and mechanics.

\section*{Conclusions}
\label{sec:conclusions}
To the best of our knowledge, we have developed the first
thermodynamically consistent transport theory applicable for pure
ILs. Our theory describes the complete set of coupled transport
equations for composition (concentration), temperature, charge
density, electric potential, and convection. We make explicit use of
the force law to include all mechanical couplings. Our detailed
equations for the electrolyte dynamics describe the individual
contributions of all species to transport of mass, charge, heat, and
to convection.

The present work completes our two-fold validation procedure. In a
first publication, the theory was validated in a joint
experimental/theoretical investigation of the double-layer-structures
formed by a binary IL near an electrified interface, and of how a
ternary salt influences these characteristic
IL-structures.\cite{C7CP08243F} Here, we apply the electroneutral
transport theory to a complete secondary battery cell.

Our simulations determine the concentration dependent electrolyte
dynamics during a complete cycle of discharging-charging of a
secondary zinc-ion battery based on an IL-electrolyte. The resulting
discharge profile is in good agreement with the experimental
results. However, the cell performance is limited by the negativity of
the zinc ions. These must overcome a Coulombic potential barrier to
maintain cell operation by the formation of concentration
gradients. In contrast to binary ILs, convection does not play a
significant role compared to the competing mechanisms of migration and
diffusion.

Our theory provides a rigorous framework for the determination of the
transport parameters, which all follow from the fundamental Onsager
matrix. Furthermore, we enrich the present discussion regarding the
interpretation of transport parameters in solvent-free electrolytes,
as they depend on different reference-frames. Our model clarifies the
exact number of independent parameters and contains simple equations
for their conversions.

The generality of the presented framework covers a huge variety of
physico-chemical systems and it can be customized by appropriate free
energy models. In particular, it applies to electroneutral
cell-systems as well as to confined geometries near electrified
interfaces. Our transport theory thus provides a valid framework for
the description of complex, multi-component battery-electrolytes in
general (concentrated electrolytes, water-in-salt-electrolytes, ILs)
as well as for the description of surface-effects of ILs and
IL-mixtures.

\section*{Acknowledgement: LuZi/BmBF}
\label{sec:ackn-luzibmbf}

This work was supported by the German Min\-istry of Education and
Research (BMBF) (project LUZI, BMBF: 03SF0499E).

% \printtables
% \printfigures

\bibliographystyle{apsrev4-1}
\bibliography{library}

\end{document}